\documentclass[traditabstract,a4,twocolumn]{aa}  

\usepackage{epsfig}
\usepackage{graphicx}
\usepackage{epstopdf}
\usepackage{color}  
\usepackage{array}   
\usepackage{units}  
\usepackage[breaklinks, colorlinks, citecolor=blue]{hyperref}
\usepackage{natbib}  
\usepackage{multirow}   
\usepackage{amsmath,amssymb}   
\usepackage[english]{babel}
\usepackage[latin1]{inputenc}
\usepackage[T1]{fontenc}
\usepackage{longtable,lscape}
\usepackage{footnote}
\usepackage{txfonts}
\usepackage[toc,page]{appendix} 

\begin{document}

\newcommand{\avg}[1]{\left< #1 \right>} 
\newcommand\red[1]{{\color{red}(\bf ?? edits: #1)}}
\newcommand\green[1]{{\color{green}(\bf MD edits: #1)}}


   \title{MILCANN : \\ A neural network assessed tSZ map for galaxy cluster detection} 

   \author{G. Hurier\inst{1}, N. Aghanim\inst{2} \& M. Douspis\inst{2}
          }

\institute{Centro de Estudios de F\'isica del Cosmos de Arag\'on (CEFCA),Plaza de San Juan, 1, planta 2, E-44001, Teruel, Spain
\and Institut d'Astrophysique Spatiale, CNRS (UMR8617) Universit\'{e} Paris-Sud 11, B\^{a}timent 121, Orsay, France\\
\\
\email{hurier.guillaume@gmail.com} 
}

   \date{Received /Accepted}
 
   \abstract{ We present the first combination of thermal Sunyaev-Zel'dovich (tSZ) map with a multi-frequency quality assessment of the sky pixels based on Artificial Neural Networks (ANN) aiming at detecting tSZ sources from sub-millimeter observations of the sky by $Planck$.  We present the construction of the resulting filtered and cleaned tSZ map, MILCANN. We show that this combination allows to significantly reduce the noise fluctuations and foreground residuals compared to standard reconstructions of tSZ maps. From the MILCANN map, we constructed a tSZ source catalogue of about 4000 sources with a purity of 90\%.  Finally, We compare this catalogue with ancillary catalogues and show that the galaxy-cluster candidates in our catalogue are essentially low-mass (down to $M_{500} = 10^{14}$ M$_\odot$) high-redshift (up to $z \leq 1$) galaxy cluster candidates. 
}


   \maketitle


\section{Introduction}

\label{sec:theory}

Galaxy clusters, being the largest virialized structures in the
Universe, are excellent tracers of the matter distribution.
Their abundance can be used to constrain the cosmological model in an independent and complementary way from Cosmic Microwave Background (CMB) \citep[see e.g.,][]{planckXXI,planck2015cosmo,hur17,sal18}.  Galaxy clusters are composed
of dark matter, stars, cold gas and dust in galaxies, and a hot
ionized intra-cluster medium (ICM).  Consequently, they can be
identified in the optical bands as concentrations of galaxies
\citep[see e.g.][]{abe89,gla05,koe07,ryk13}, they can be observed in
X-rays thanks to the bremsstrahlung emission produced by the ionized
intra-cluster medium (ICM) \citep[see
  e.g.][]{boh00,ebe00,ebe01,boh01}. The same hot ICM also creates a
distortion in the black-body spectrum of the CMB through the thermal Sunyaev-Zel'dovich (tSZ) effect
\citep{sun69,sun72}, an inverse-Compton scattering between the CMB photons and the ionized electrons in the ICM, \citep[see e.g.][for reviews]{bir99,car02}. The  thermal SZ Compton parameter in a given direction, $\vec{n}$, on the sky is given by
\begin{equation}
y (\vec{n}) = \int n_{e} \frac{k_{\rm{B}} T_{\rm{e}}}{m_{\rm{e}} c^{2}
} \sigma_{T} \ \rm{d}s
\label{comppar}
\end{equation}
where d$s$ is the distance along the line-of-sight, $\vec{n}$, and $n_{\rm{e}}$
and $T_{e}$ are the electron number density and temperature,
respectively.
In units of CMB temperature the contribution of the tSZ effect at a frequency $\nu$ is
\begin{equation}
\frac{\Delta T_{\rm{CMB}}}{T_{\rm{CMB}} }= g(\nu) \ y.
\end{equation}
Neglecting relativistic corrections, we have 
\begin{equation}
g(\nu) = \left[ x\coth \left(\frac{x}{2}\right) - 4 \right],
\label{szspec}
\end{equation}
with $ x=h \nu/(k_{\rm{B}} T_{\rm{CMB}})$. At $z=0$, where $T_{\rm
  CMB}(z=0)$~=~2.726$\pm$0.001~K, the tSZ effect is negative below
217~GHz and positive for higher frequencies.\\

Recent large cluster catalogues based on measurements of the tSZ effect have been produced from {\it $Planck$} \citep{planckESZ,planckPSZ}, ACT \citep{mar11,act18}, and SPT \citep{ble14} data. Several
detection algorithm targeted to tSZ sources \citep[see e.g.,][]{mel06,car09} have been
proposed and compared \citep{mel12}. 
This demonstrated that a
multi-filter approach based on the use of optimal filters for the tSZ detection is more robust than a
tSZ map-based approach. The latter relies on the construction of $y$ map with component separation methods such as MILCA \citep{hur13} or NILC \citep{rem11}. These methods are devised to mitigate the contamination of the tSZ signal by
other astrophysical emissions, mainly radio, infra-red point sources, and cosmic infra-red background
\citep{dun11,shi11,rei12,sie13,planckSZS}. The combination of high and low resolution SZ surveys, as was performed in PACT \citep{pact}, is another way of efficiently reducing the contamination of the $y$ map but this relies on the availability of complementary public data. In all cases, the reconstructed $y$ map cannot be totally immune from contamination that can produce spurious galaxy cluster detections and/or a significant bias the measured tSZ fluxes. \cite{mel12} showed that tSZ map-based detection methods imply a larger number of spurious tSZ sources than multi-filter methods, and hence a significantly lower level of purity of the produced catalogues. As a result, an a-posteriori quality assessment of the tSZ signal from galaxy clusters is required to produce high-purity galaxy cluster samples of tSZ clusters.\\
In \citet{planckPSZ} and \citet{PSZ2}, the quality assessment of the tSZ sources, named validation, was detailed. It included several steps from the cross-matches with catalogues to the search for counterparts in galaxy or X-ray surveys, including visual inspection of tSZ sources. Automatic assessments of the quality of tSZ detected sources can also be performed. A method based on artificial neural networks (ANN) was proposed by \cite{agh14}. It uses the {\it $Planck$} multi-frequency data to assess the quality of the tSZ sources by decomposing the measured signal into the different astrophysical components contributing to the {\it $Planck$} frequencies. This new quality assessment method was applied to validate the {\it $Planck$} cluster catalogue \citep{PSZ2}. Moreover, the use of the ANN method showed that the tSZ-source catalogue \citep{planckPSZ} suffers from contamination by galactic CO sources and infra-red emission. Recent follow-up of $Planck$ tSZ sources in the optical \citep{van16}, have shown the efficiency of this ANN-based quality assessment by confirming the spurious nature of tSZ sources with bad quality criteria.
\\

In this work, we extend the use of the ANN method initially developed by \cite{agh14} to provide a quality assessment of individual tSZ sources to provide a quality assessment of each pixel in a reconstructed full-sky $y$ map. The resulting ANN-weighed $y$ map used then for clusters detection can provide a new catalogue of tSZ sources. The paper is organised as follows. In section~\ref{sec_data}, we present the different datasets. In section~\ref{ann_sec}, we detail the construction of the ANN-based weights. In
section~\ref{carac_sec}, we construct a sample a galaxy cluster
candidates and we present a detailed characterization of this sample. Finally in sect.~\ref{sec_probe}, we perform a multi-wavelength assessment of the detected galaxy cluster candidates.\\

Throughout the paper we use the following cosmological parameters, $\Omega_{\rm m} = 0.316$, $H_0 = 67.26$, $\sigma_8 = 0.83$, $n_s = 0.9652$, and $\Omega_{\rm b}h^2 = 0.02222$ derived from $Planck$ collaboration 2015 results \citep{planck2015cosmo}.

\section{Data}
\label{sec_data}

For the present analysis we used several publicly available datasets (catalogues and surveys) either to describe astrophysical source properties or to characterize the galaxy cluster candidates detected in the ANN-weighted $y$ maps.

We use the {\it $Planck$} intensity maps from 70 to 857~GHz \citep{PlanckMIS}. We also use the spectral responses given in \citet{planckRESP}. We assume the {\it $Planck$} beams are Gaussian with values given in \citep{planckBEAM}. We also use the {\it $Planck$} full-sky CMB-lensing map \citep{plcklens}. The {\it $Planck$} collaboration has published source catalogues that we use for our analysis. Namely, the catalogue of tSZ sources detected in {\it $Planck$} data \citep[PSZ2 hereafter, see ][]{PSZ2}, and the catalogues of point-sources detected at 30~GHz and 353~GHz from the $Planck$ Catalogue of Compact Sources \citep{planck2013-p05}. All the {\it $Planck$} data can be retrieved from the $Planck$ Legacy Archive\footnote{\url{https://pla.esac.esa.int/}}

We also use the reprocessed IRAS maps, IRIS \citep[Improved Reprocessing of the IRAS Survey,][]{miv05}, the AllWISE Source Catalog\footnote{\url{http://wise2.ipac.caltech.edu/docs/release/allwise/expsup/}} \citep{wri10,mai11}. Finally, we use catalogues of clusters detected in the X-rays \citep[MCXC,][and reference therein]{pif11} and in the SDSS survey namely WHL12 \citep{wen12}, WHL15 \citep{wen15}, WHY18 \citep{wen18} and redMaPPer \citep{ryk13}.

\section{Artificial neural network}

\label{ann_sec}
Machine learning, and in our specific case the ANN, enables to directly learn from the data the characteristic signature of "true" tSZ sources and spurious signal using a reference sample of astrophysical sources. As shown in \cite{agh14},\cite{van16} and \cite{planckPSZ,PSZ2}, this method allows to identify spurious tSZ sources from catalogues of cluster candidates.
In the following, we adapt the approach used in \cite{agh14} in order to extend the machine-learning based quality assessment to each pixel of the sky-maps rather than to samples of individual detected tSZ candidates.\\

For clarity, we first summarize here the key elements of the ANN method (a more detailed description is provided in \cite{agh14}).

\begin{figure}[!th]
\center
\includegraphics[width=8cm]{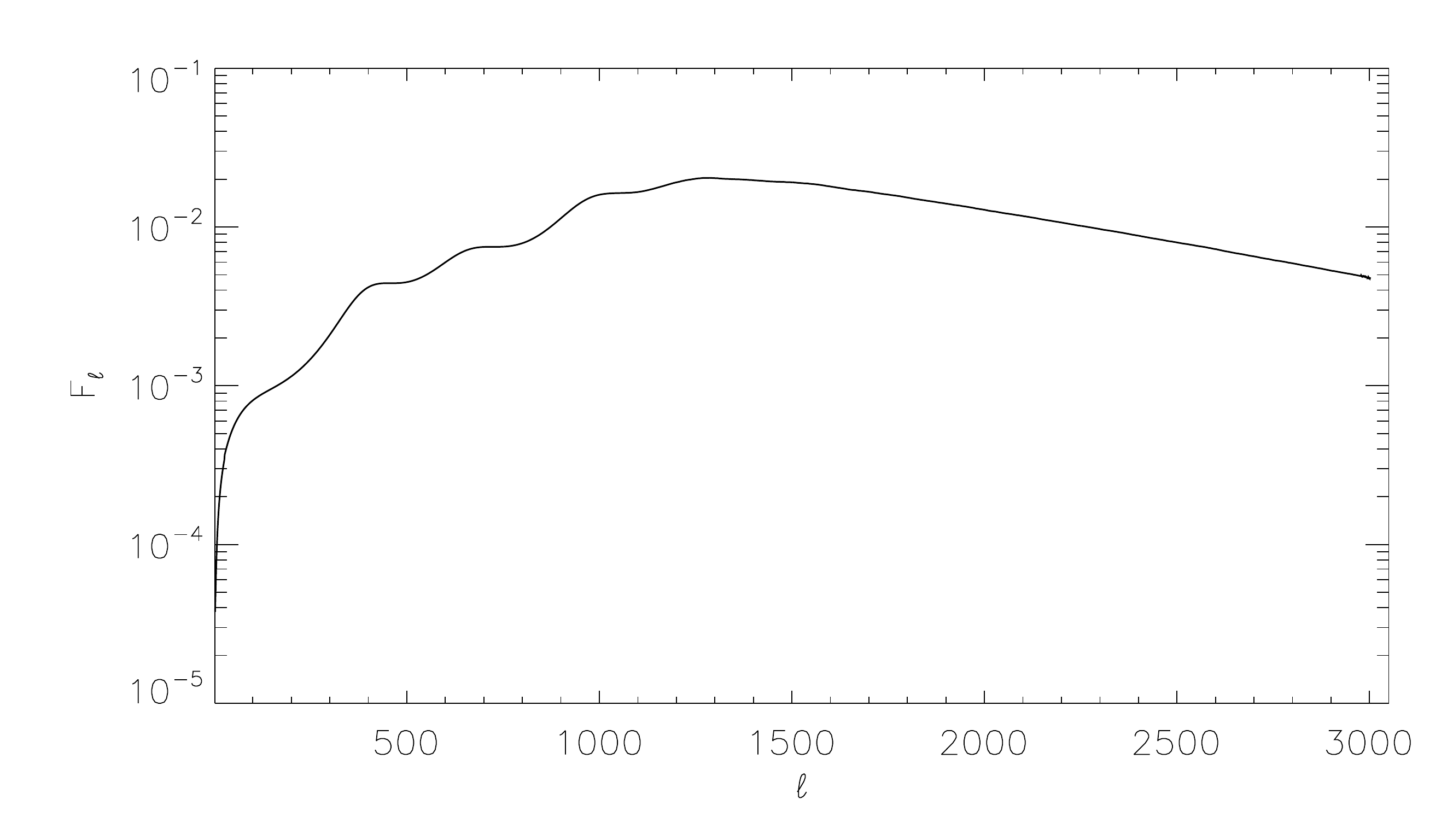}
\caption{Matched filter in $\ell$ space ($F_\ell$) used for the ANN-based quality assessment.}
\label{fl}
\end{figure}

We focus on the astrophysical emissions that
affect the most the tSZ signal in multifrequency experiments and we model the flux at each frequency taking into account tSZ effect (neglecting relativistic
corrections), CMB, and CO emission. We also add an
effective IR component standing for the contamination by galactic dust, cold Galactic sources, and CIB fluctuations; and an
effective radio component accounting for diffuse radio and synchrotron
emission and radio sources. The flux in frequency is then written as
\begin{align}
F_\nu =& A_{\rm SZ} F_{\rm SZ}(\nu) + A_{\rm CMB} F_{\rm CMB}(\nu) +
A_{\rm IR} F_{\rm IR}(\nu) \nonumber \\
&+ A_{\rm RAD} F_{\rm RAD}(\nu) + A_{\rm CO}
F_{\rm CO}(\nu) + N(\nu),
\label{eq:model}
\end{align}
where $F_{\rm SZ}(\nu)$, $F_{\rm CMB}(\nu)$, $F_{\rm IR}(\nu)$,
$F_{\rm RAD}(\nu)$, and $F_{\rm CO}(\nu)$ are the spectra of tSZ, CMB,
IR, radio, and CO emissions; $A_{\rm SZ}$, $A_{\rm CMB}$, $A_{\rm
  IR}$, $A_{\rm RAD}$, and $A_{\rm CO}$ are the corresponding
amplitudes; and $N(\nu)$ is the instrumental noise.\\

\begin{figure}[!th]
\center
\includegraphics[width=9cm]{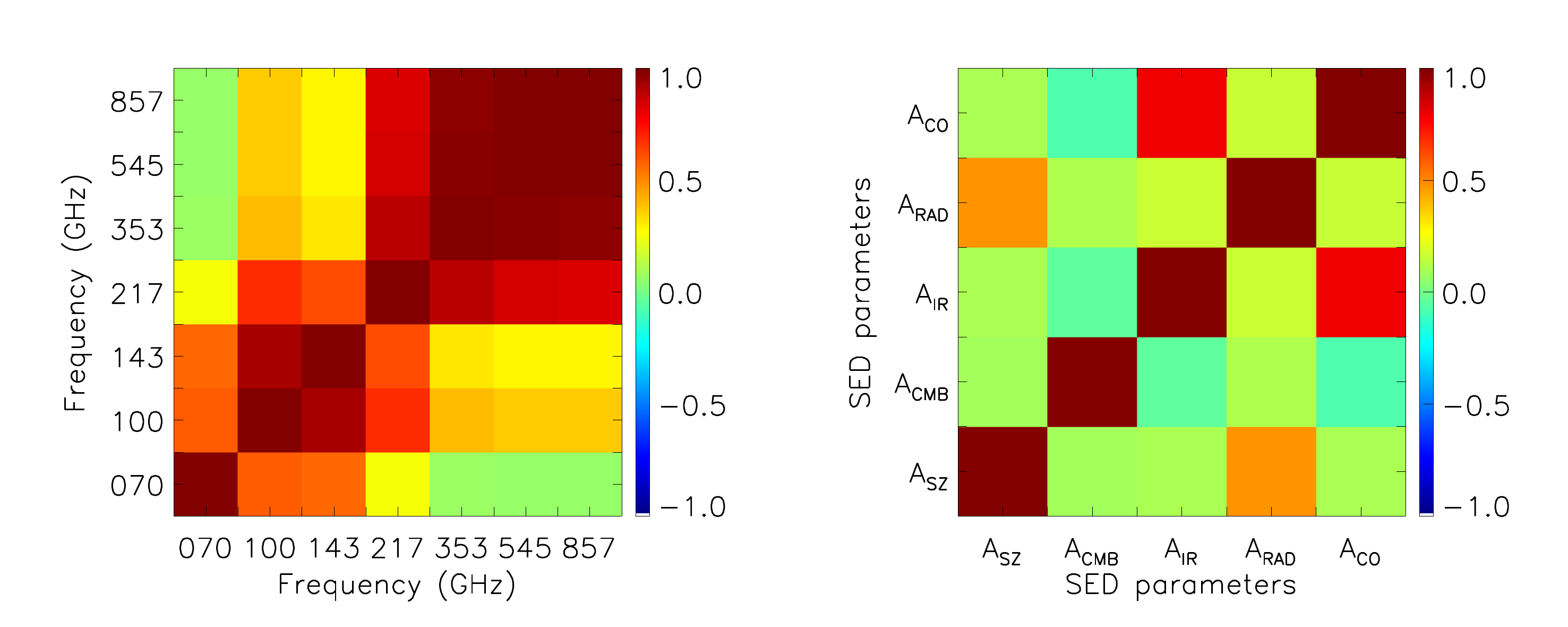}
\caption{Left panel: correlation matrix of the measured fluxes from
  30 to 857~GHz estimated on 2000 random positions over the sky. Right
  panel: correlation matrix of fitted SED parameters from the same
  positions.}
\label{cormat}
\end{figure}

In order to improve the photometry of tSZ sources, we compute the flux for each pixel and each frequency using a matched-filter in $\ell$ rather than the aperture photometry used in \citet{agh14}. To build the matched-filter in $\ell$ equivalent to a  Wiener filter, we compute the power spectrum, $y_\ell$, of a tSZ signal from a single cluster with $R_{500} = 5'$ (point-like with respect to the $Planck$ experiment) assuming a universal pressure profile \citep{arn10}. We also consider the power spectrum of the CMB, $C^{CMB}_\ell$, computed using $Planck$ best-fit cosmology \citep{planck2015cosmo}, and the power spectrum of the noise, $C_\ell^{\rm NN}$, in the 100~GHz channel (estimated from the half-ring map difference). Considering the most relevant frequencies for the tSZ flux estimation (100 to 217 GHz), relevant angular scales ($\ell \in [1000,2000]$) and focusing on the cleanest sky area (high galactic latitudes, $b > 20°$), we neglect the thermal dust contribution from the Milky Way and hence do not include it in the computation of the matched-filter. The resulting filter is thus defined as:
\begin{align}
F_\ell = \sqrt{\frac{y_\ell}{C^{\rm CMB}_\ell + C_\ell^{\rm NN}}}.
\end{align}
It is shown in Fig.~\ref{fl}. The filter is applied to the $Planck$ intensity maps from 70 to 857 GHz convolved with a 13' beam (lowest resolution associated with the 70~GHz map). The resulting harmonic space coefficients are thus given by $a'_{\ell, m} = F_\ell a_{\ell, m}$. The amplitudes of $F_\nu$ (Eq. \ref{eq:model}), namely $A_{\rm SZ}$, $A_{\rm CMB}$,
$A_{\rm IR}$, $A_{\rm RAD}$, and $A_{\rm CO}$, are linearly fitted for each pixel from the map constructed with the filtered harmonic coefficients. We verified that the results do not significantly depend on the chosen amplitude of $C_\ell^{\rm NN}$ for the matched-filter computation. Figure~\ref{cormat} (right panel) shows the correlation matrix of the fitted SED parameters. The left panel displays the correlation matrix of measured fluxes (via the matched-filter) from 70 to 857~GHz. The correlation matrix of the SED amplitudes shows that the matched-filter photometry allows to achieve a higher signal-to-noise measurement of the fluxes and thus a better separation of the various contributions to the SED, as compared to the aperture photometry used in \citet{agh14}.
The difference is particularly striking for the tSZ component that is now only significantly correlated with the radio component. This correlation has two origins, a physically-motivated spatial correlation with radio-loud AGNs and the similarities of shape between the tSZ and radio SED at low frequency, making their distinction hard to achieve. We observe another significant correlation between the CO and the thermal dust components due to spatial correlation. We stress that these two components are among the major sources of spurious detection in the {\it $Planck$} catalogue: the CO emission produces a rise of the intensity at 100 GHz, and the dust emission an increasing signal with frequency. These two trends together mimic the tSZ spectral signature. In this case, the 70 GHz channels is of great use to separate CO emission, only affecting 100 GHz channels, from a tSZ emission that should present consistent 70 and 100 GHz channels.
Additionally by construction, the matched-filter selects specific scales of the tSZ emission corresponding to the cluster scales allowing to reduce the contamination by large scale emissions (i.e., galactic thermal dust, CMB).

\begin{figure*}[!th]
\center
\includegraphics[width=0.45\linewidth]{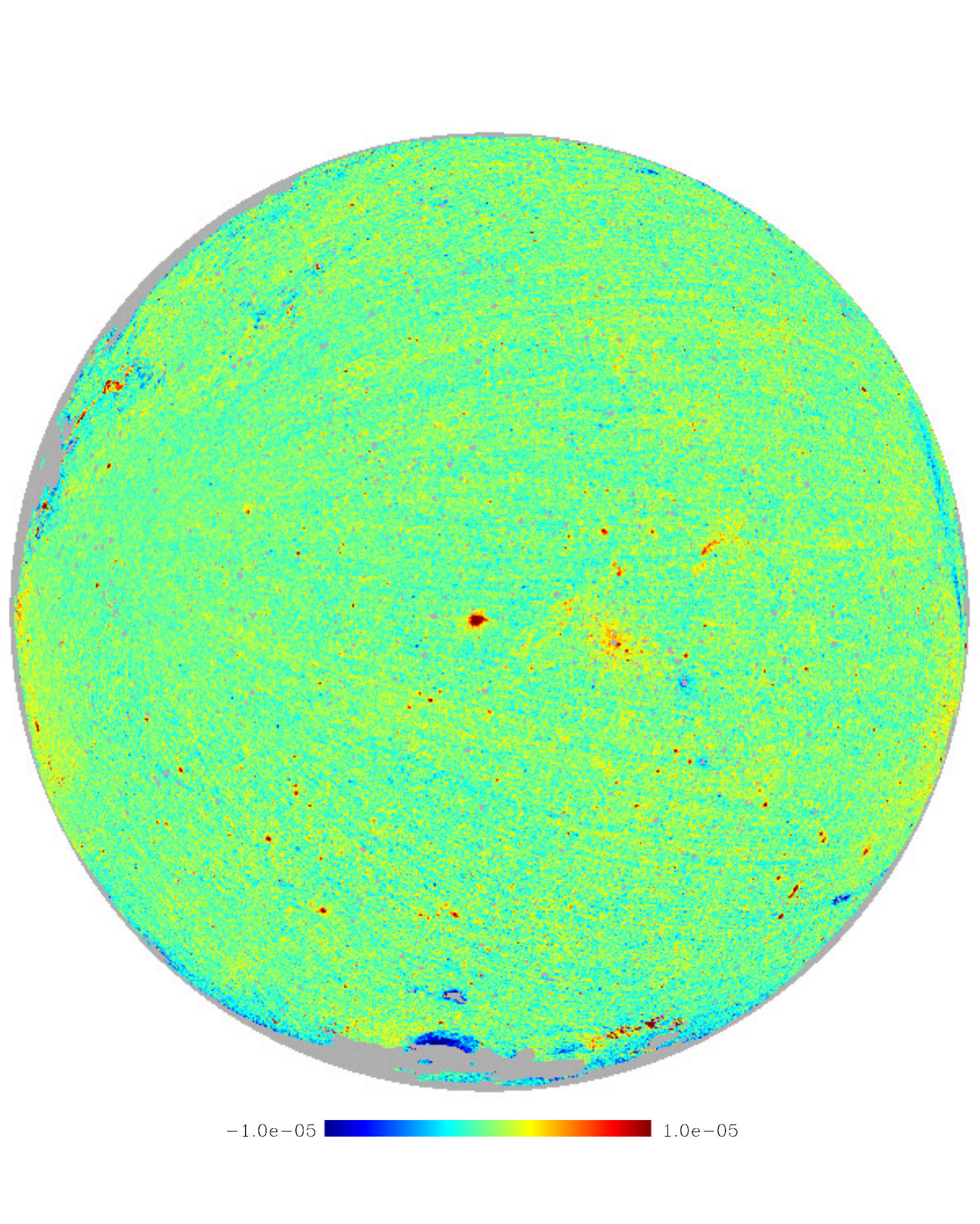}
\includegraphics[width=0.45\linewidth]{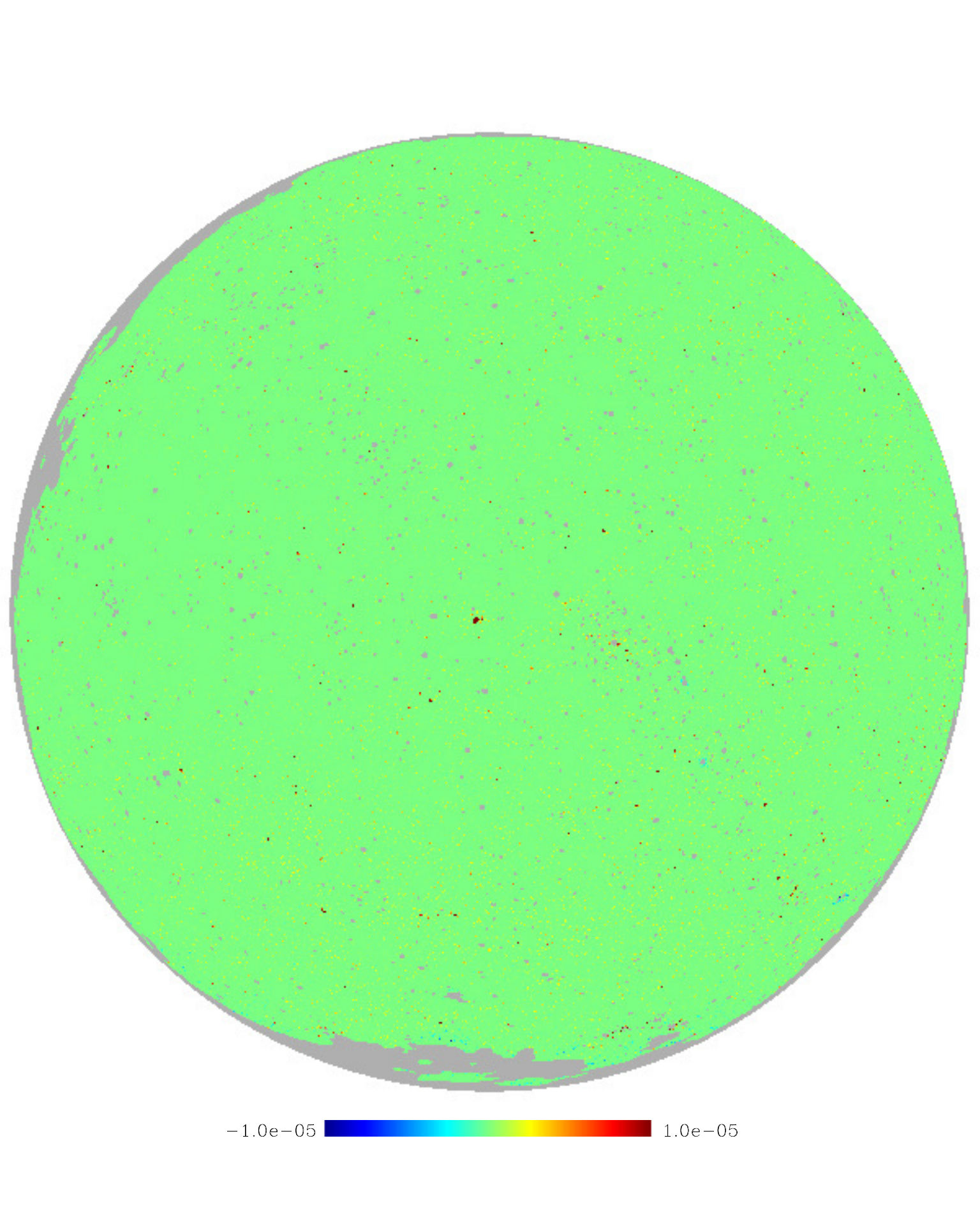}\\
\caption{Left panel: $Planck$ tSZ MILCA map. Right panel: MILCANN map, for the northern-sky in orthographic projection. Gray regions are masked due to galactic-foreground or point-source contamination.}
\label{fullsignal}
\end{figure*}

\begin{figure*}[!th]
\center
\includegraphics[width=0.9\linewidth]{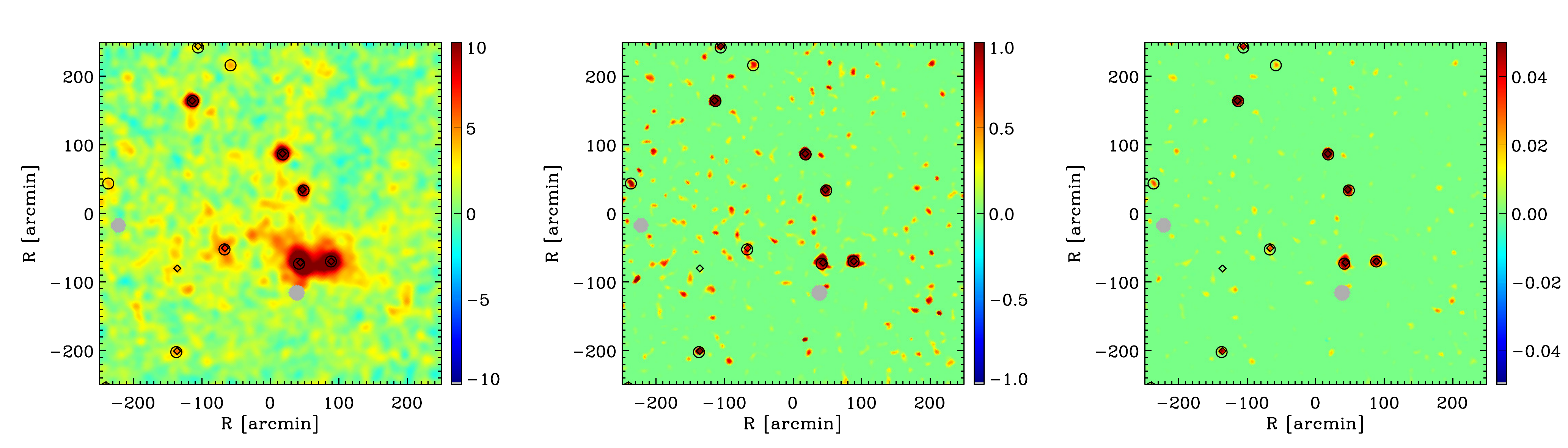}
\includegraphics[width=0.9\linewidth]{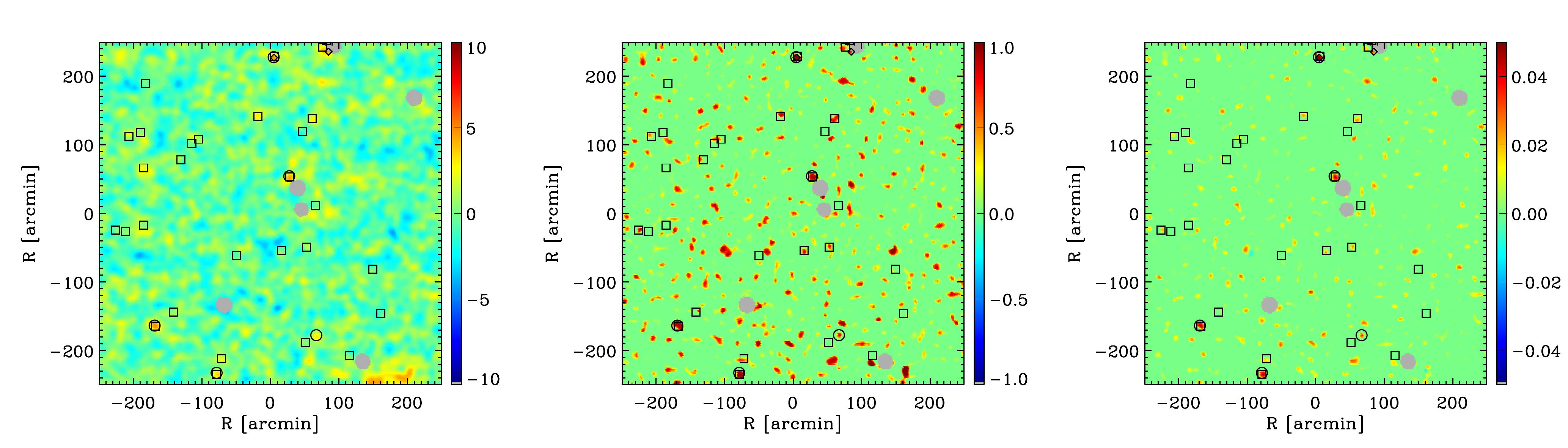}
\caption{ Upper panels show a region of the sky including a nearby multiple-cluster system, whereas lower panels show a region with a large number of clusters identified in the optical within SDSS data.
From left to right: $Planck$ tSZ MILCA map, neural network weight, match-filtered MILCANN map, for patches of 8.5 $\times$ 8.5 degrees in gnomonic projection centered on galactic coordinates (l,b) = (264$^{\rm o}$,-24$^{\rm o}$) and (l,b) = (11.5$^{\rm o}$,70$^{\rm o}$). Grey regions are masked due to point-source contamination. We display objects from the HAD catalogue (presented in this paper) as black circles, from the PSZ2 catalogue as black diamond, and from redMaPPer (when available) as black squares.
}
\label{patchsignal}
\end{figure*}

\begin{figure*}[!th]
\center
\includegraphics[width=0.9\linewidth]{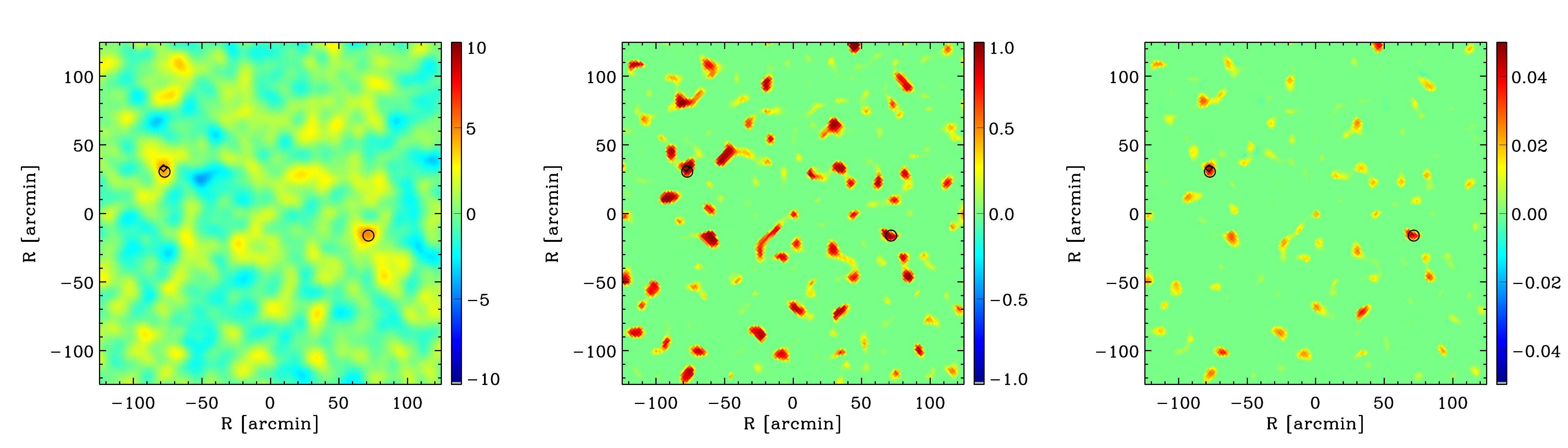}
\includegraphics[width=0.9\linewidth]{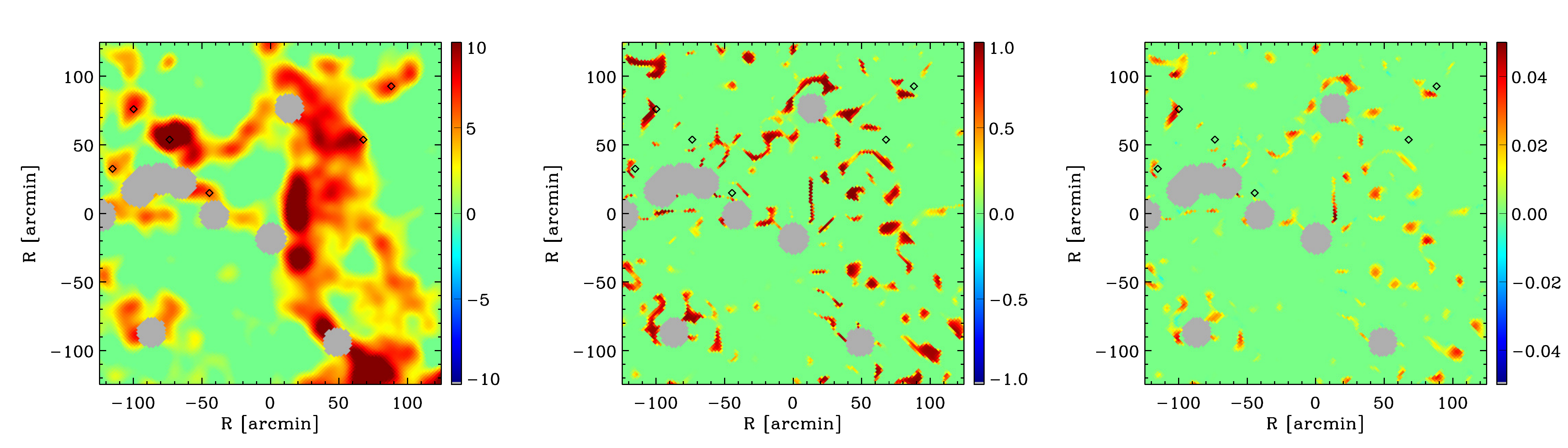}
\caption{ Upper panels show a region of the sky without foreground contamination, whereas lower panels show an area of the sky dominated by galactic contamination. From left to right same
 as Fig.~\ref{patchsignal} for patches of 4.25 $\times$ 4.25 degrees centered on galactic coordinates (l,b) = (44.5$^{\rm o}$,-27.5$^{\rm o}$) and (l,b) = (122$^{\rm o}$,24.5$^{\rm o}$)}
\label{patchsignal2}
\end{figure*}
Following \citet{agh14}, we consider a standard three-layer back-propagation ANN to separate
pixels of the sky maps into three populations of reliable quality, unreliable quality that is false detection, and noisy sources \citep[they are referred to as good bad and ugly in][]{agh14}. The inputs of the neural network consist of
the five SED parameters per pixel, computed here with a  more precise photometry based on the matched-filter, from which we derive three full-sky maps associated with the three quality classes.\\

To train the neural network and to assess the tSZ signal quality for each sky-pixel, we use the same sample for {\it Good}, {\it Bad}, and {\it Ugly} classes as in \citet{agh14}, namely: galaxy clusters; infra-red, radio, and cold galactic sources; and random (noise) estimates.

\section{Construction of the tSZ map}
\subsection{MILCA $Planck$ tSZ map}

Independently from the ANN quality assessment, we reconstructed a tSZ map with the MILCA method
\citep{hur13}, using $Planck$ HFI from at 100 to 857 GHz, after verifying
that including frequencies from 30 to 70 GHz does not change significantly the reconstructed map, especially at galaxy cluster
scales.\\ We performed the construction of the tSZ map using 8 bins
in spherical harmonic space. For the first three bins, we used two constraints (tSZ and CMB), and for the last five bins we only used a constraint on the tSZ SED. The map reconstruction was performed with an effective FWHM of 7 arcmin. For all bins, two degrees of freedom were used to minimize the variance of the noise \citep[see][for a detailed description of the MILCA method]{hur13}.\\ 

In Figure ~\ref{fullsignal}, we show the MILCA full-sky map at 7 arcmin FWHM.
Figures~\ref{patchsignal}~and~\ref{patchsignal2} show a zoom on two regions of $8.5
  \times 8.5$ degrees and two regions of $4.25
  \times 4.25$ degrees where we can observe bright galaxy clusters; a typical region with Planck clusters, a region with known optically selected clusters (redMaPPer catalogue with $\lambda > 50$); a region showing low-S/N cluster candidates, and a region showing a significant CO contamination triggering spurious detection in the PSZ2 catalogue.
In the full-sky map (Fig. ~\ref{fullsignal}), we observe a significant amount of foreground
residuals near the galactic plane, where synchrotron and free-free
residuals appear as negative biases in the tSZ $y$-map signal. We also
observe contamination by bright galactic cirrus correlated with the
zodiacal light. As shown in previous works \citep{hur13,planckszmap}, the main sources of contamination in tSZ maps built from $Planck$ intensity maps are radio point sources, CO, and CIB
emission.  
\begin{figure}[!th]
\begin{center}
\includegraphics[scale=0.2]{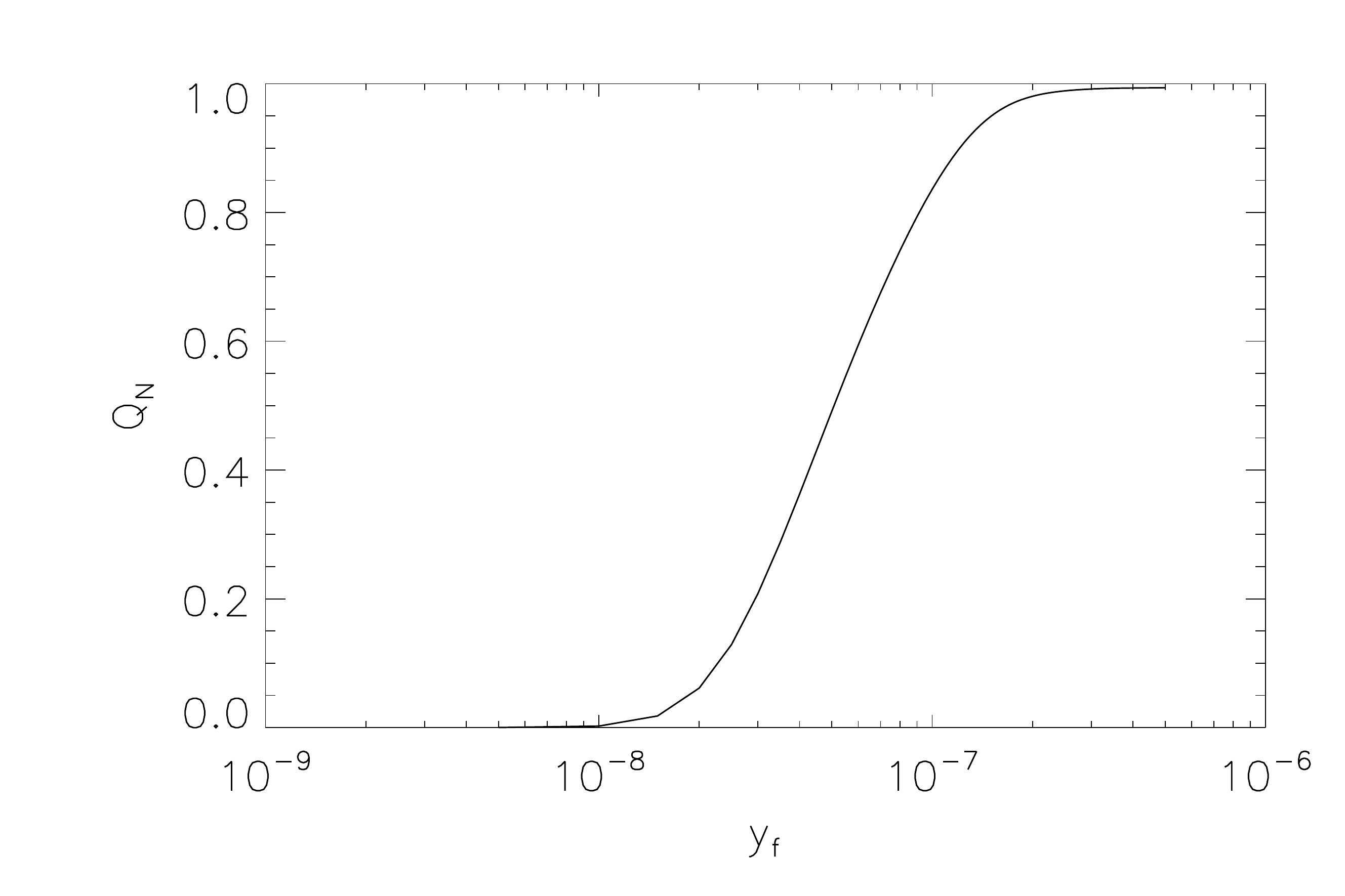}
\caption{ANN-weight average value as a function of the match-filtered intensity.}
\label{aresp}
\end{center}
\end{figure}

\begin{figure*}[!th]
\center
\includegraphics[width=0.45\linewidth]{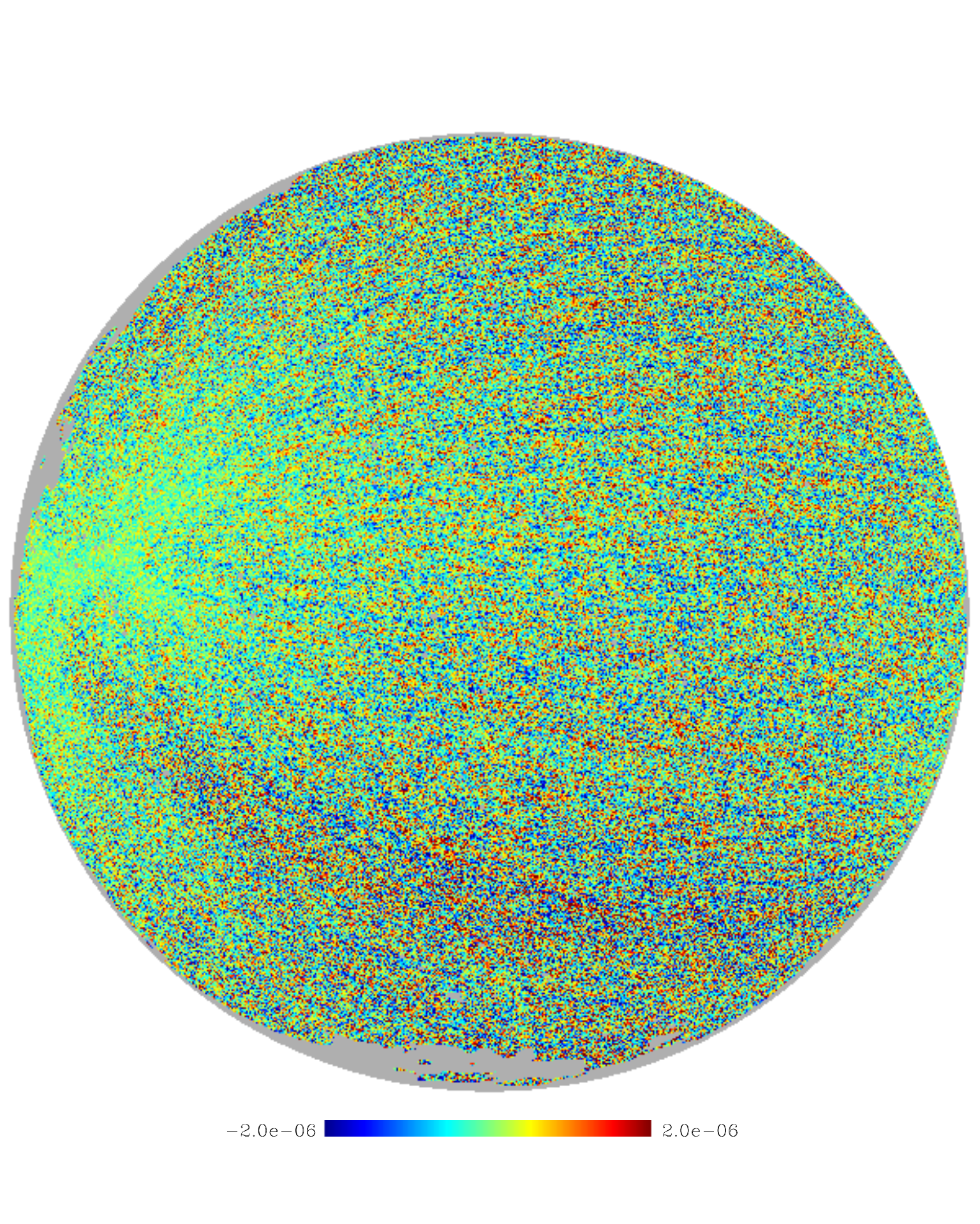}
\includegraphics[width=0.45\linewidth]{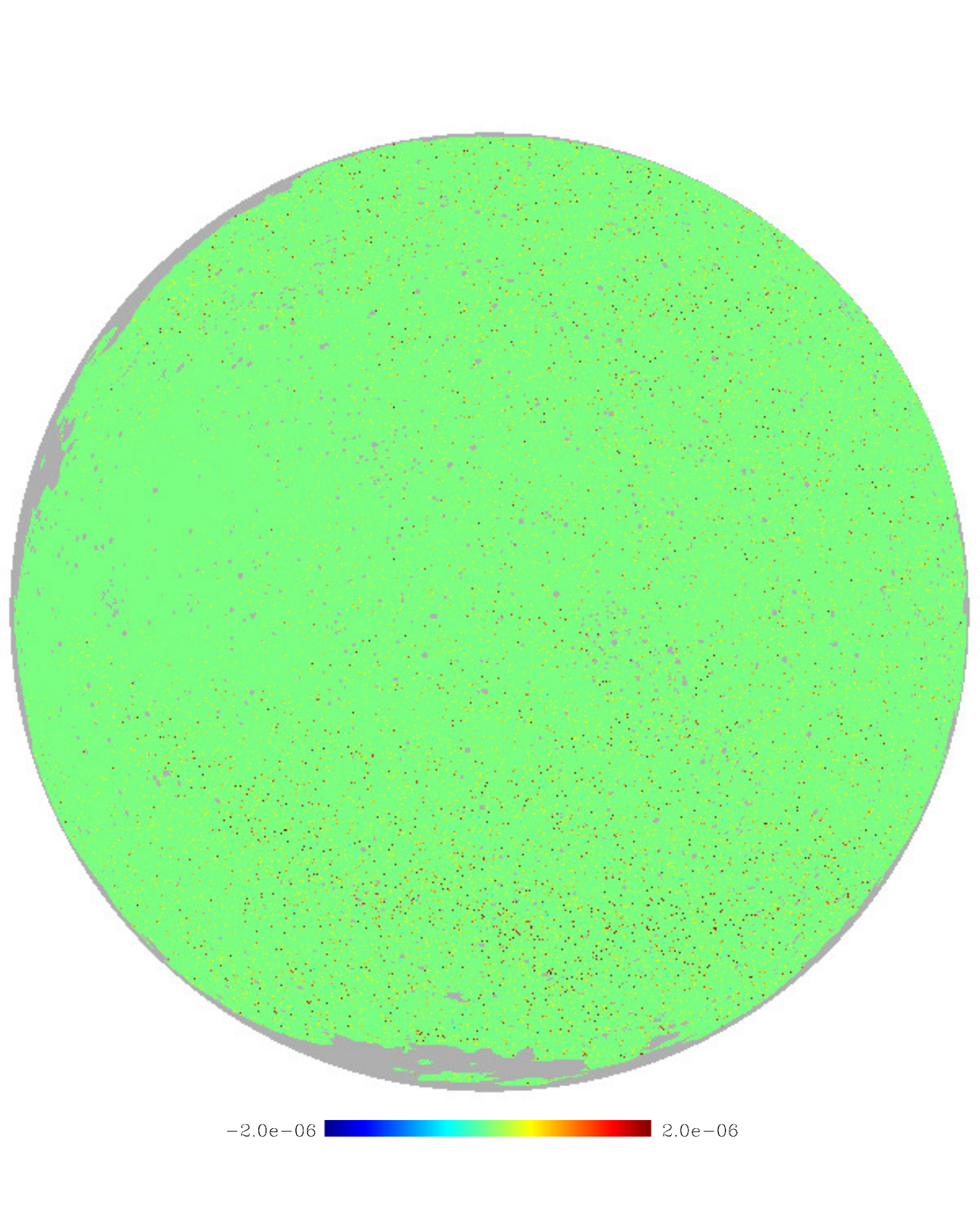}
\caption{Left panel: simulated $Planck$ tSZ MILCA noise map. Right panel: simulated MILCANN noise map, for the northern-sky in orthographic projection. Grey regions are masked due to galactic foregrounds or point-source contamination.}
\label{fullnoise}
\end{figure*}

For consistency with the ANN quality-assessment, we convolve the reconstructed tSZ map,
noted $\widehat{y}$, by the matched-filter used for the SED fitting. The obtained filtered map, $\widehat{y}_f$, has a transfer function consistent with the maps used to perform the ANN classification.\\

\subsection{ANN weighting}

In Figure ~\ref{fullsignal}, we observed that the reconstructed tSZ map 
suffers from bias due to residuals from other astrophysical emission.  Using the neural-network based quality
assessment presented in sect.~\ref{ann_sec}, we can estimate
the quality of the tSZ signal for each line-of-sight in the sky.
First, we define an ANN-based weight, $Q_{\rm N}$, as
\begin{align}
Q_{\rm N} = Q_{\rm GOOD} (1-Q_{\rm BAD}),
\end{align}
where $Q_{\rm GOOD}$ and $Q_{\rm BAD}$ are the ANN classification
output values for the {\it Good} and {\it Bad} classes. By
construction, this ANN-weight ranges from 0 to 1, with
values close to 1 for pixels that present a high-quality tSZ signal. As verified by the optical follow-up of tSZ candidates in \citet{van16}, this ANN-based weight provides a good proxy for the robustness of a tSZ signal.\\

\subsection{MILCANN map}

Finally, we construct from the input tSZ map a new map (noted MILCANN) filtered and cleaned from the contamination as:
\begin{align}
M_{\rm MILCANN} = \widehat{y}_f \, Q_{\rm N}.
\end{align}
In Figure ~\ref{patchsignal}~and~\ref{patchsignal2}, we display the
ANN weight described in Sect.~\ref{ann_sec} and the MILCANN map. We
observe a clear spatial correlation between the ANN weight and the input tSZ map. 
In Fig.~\ref{fullsignal}, the MILCANN map shows that the ANN weight
has significantly removed the foreground contamination, especially
near the galactic plane where synchrotron and free-free contamination
have been completely suppressed. Similarly, the contamination produced
by high-latitude galactic dust has been reduced by the ANN-weighting procedure.  On
Figs.~\ref{patchsignal}~and~\ref{patchsignal2}, we also observe that the MILCANN map presents a significantly reduced overall background noise, with the bright tSZ pixels conserved. \\
We observe that redMaPPer clusters previously undetected via their tSZ signal are seen in the MILCANN map (see second row of Fig.~\ref{patchsignal}). We also note that the ANN-weighting process allows us to avoid the contamination by spurious CO sources that were present in the PSZ2 catalogue (see bottom row of Fig.~\ref{patchsignal2}). 
In general, Figs.~\ref{patchsignal}~and~\ref{patchsignal2} show that the ANN-weighting process leading to the MILCANN map is a significant improvement for galaxy cluster detection when approaching the tSZ noise limit. It allows us to lower the threshold of detection without having a significant amount of spurious sources.
\\

However, it is worth noting that the ANN classification, and subsequent weighting process is not a perfect procedure  \citep[see also][]{agh14}. As a matter of fact, the ANN-weight map in Fig.~\ref{patchsignal} presents values close to 1 even for some pixels where no clear tSZ signal can be
observed. These misclassified pixels are produced by chance alignment
between noise structure and tSZ spectral signature. As a consequence,
a tSZ-source detection performed directly on the ANN-weighted map would
lead to false detections.\\ 
Furthermore, the value of $Q_{\rm N}$ is not exactly 1 for all bona fide galaxy clusters. A high-flux galaxy cluster will have $Q_{\rm N} \simeq 1$, whereas low-flux galaxy clusters will have $Q_{\rm N} < 1$. As a results, the ANN weighting process does
not conserve the shape of the tSZ sources. It modifies the intensity of faint tSZ pixels in the outskirts of galaxy clusters.\\
To estimate the transfer function of the ANN-weighting procedure, we randomly selected 10,000 pixels with no significant astrophysical emission within the 84\% sky-area used for the detection \citep[see][for a description of the mask]{planckPSZ}. In the filtered frequency maps, we added a given tSZ signal to the corresponding pixels.
We then computed $Q_{\rm N}$ for these 10,000 pixels, by applying the ANN to the modified frequency map pixels, and averaged them as a function of the injected tSZ signal. This approach allows us to properly account for real sky background and noise level in the transfer function.\\
In Figure ~\ref{aresp}, we show
the average value of the ANN-weight as a function of the injected tSZ
signal intensity after filtering, $y_f$. We observe that the ANN response presents a steep
transition, all signal below $y_f = 10^{-8}$ is completely suppressed
by the ANN-weighting process, whereas signal above $y_f = 2\, 10^{-7}$ is almost not
affected. We stress here that these Compton $y$ values are obtained after filtering and are not comparable with the tSZ intensity in the input $y$ map.  

\section{Uncertainty and systematics in the MILCANN map}
\label{sec_noise}

\subsection{Noise and CIB-residual simulations}

 We have shown qualitatively that the MILCANN tSZ map
  presents a significantly reduced background compared to the input MILCA tSZ
  map. In this section, we describe our modeling of the noise and
  CIB-residuals in the MILCA and MILCANN tSZ maps to effectively quantify the
  improvement obtained by the ANN-weighting process.\\

The tSZ maps, derived from component separation methods, are
constructed through linear combination of $Planck$ frequency maps
that depends on the angular scale and the pixel, $p$, as
\begin{align}
\widehat{y} = \sum_{i,\nu} w_{i,p}(\nu) T_{i,p}(\nu),
\end{align}
$T_{i,p}(\nu)$ is the {\it $Planck$} map at frequency $\nu$ for the
angular filter $i$, and $w_{i,p}(\nu)$ are the
weights of the linear combination (MILCA weights in this study).  The CIB contamination (or leakage) in the
$y$-map reads,
\begin{align}
y_{\rm CIB} = \sum_{i,\nu} w_{i,p}(\nu) T^{\rm CIB}_{i,p}(\nu),
\end{align}
where $T^{\rm CIB}(\nu)$ is the CIB emission at frequency $\nu$.
Using the weights $w_{i,p}(\nu)$, and considering the CIB luminosity
function, it is possible to compute the expected CIB leakage as a
function of redshift by propagating the SED through
the weights used to build the tSZ map.  As shown by
\citet{planckszcib}, the CIB at low-$z$ leaks with a small amplitude
in the tSZ map, whereas high-$z$ CIB produces a higher, dominant,
level of leakage. 

The CIB power spectra have been constrained in previous
analyses \citep[see e.g.,][]{planckcib}. They can be used to predict
the expected CIB leakage. To do so, we
performed 200 Monte-Carlo simulations of multi-frequency CIB maps that follow the CIB auto- and cross-power spectra. Then, we added instrumental noise to the simulated CIB maps.  Finally, we
applied  to these simulations the MILCA weights used to build the input tSZ map. We
obtained 200 realizations of instrumental noise and CIB in the MILCA tSZ map, consistent with noise and CIB observed in
the $Planck$ frequency maps.

It is important to stress that the noise in the input MILCA tSZ map is by construction correlated with the noise in the frequency maps. Hence, the noise on the ANN weights is also correlated with the noise in the MILCA tSZ map.  Consequently, to produce a fair description of
the noise, we trained other neural networks on the simulated
maps to reproduce the correlation feature between the noise in MILCA
map and the noise in the ANN weights.  
For completeness and considering that the training of a neural network is a non-linear process, we also added CMB, point
sources, and thermal dust to the noise+CIB simulations during the
training process.\\ Finally, we built and applied this noise-based ANN weights
to the MILCA noise+CIB-residuals simulation.  In figure.~\ref{fullnoise},
we present a simulation of noise+CIB residuals before and after applying
the noise-based ANN weights. We observe that the weighting process allows us
to significantly reduce the noise level in the simulated MILCA map.

\begin{figure}[!th]
\center
\includegraphics[width=8cm]{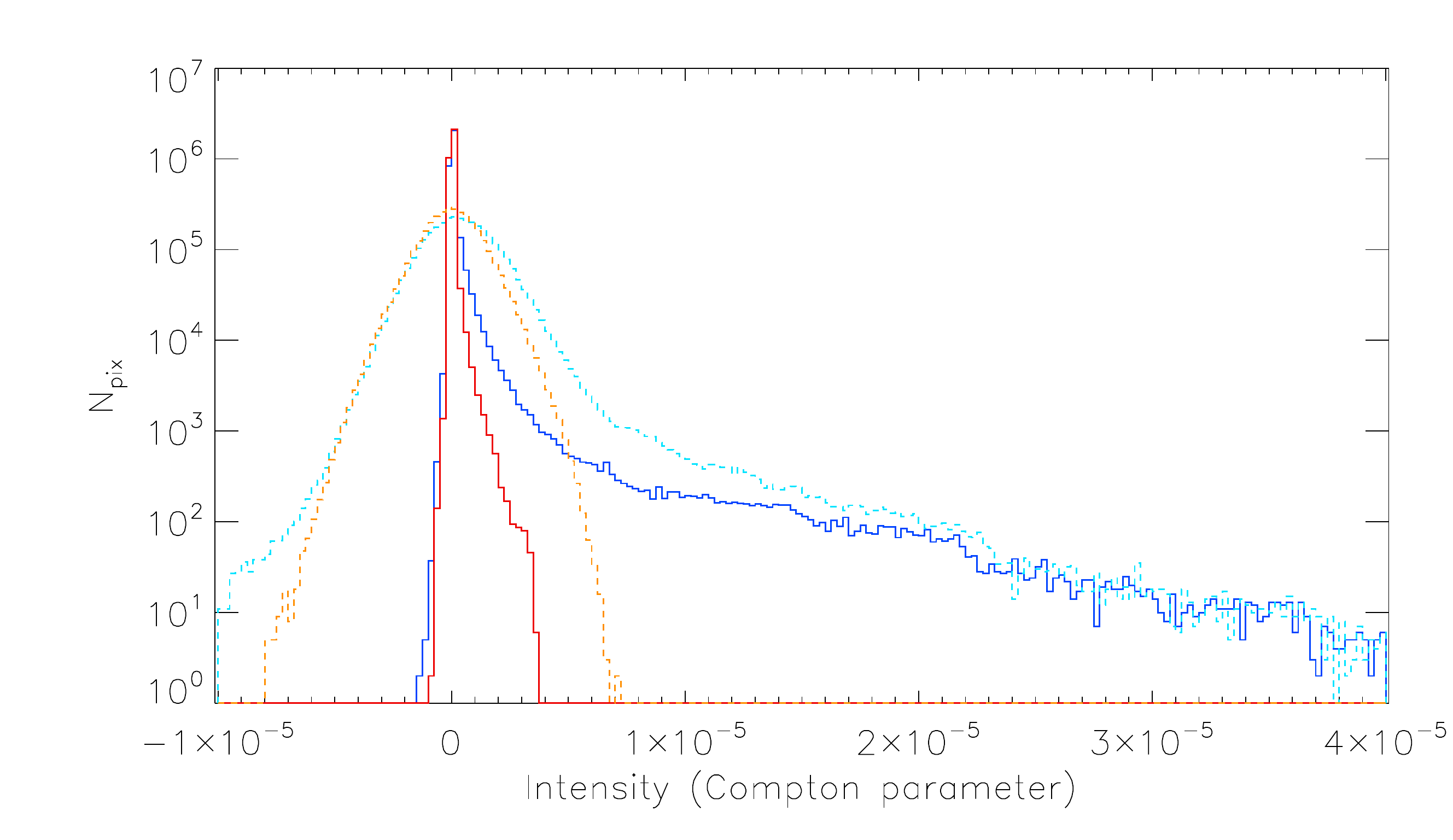}
\caption{Intensity distribution of pixels in the tSZ maps for: the MILCA map (light
  blue), the MILCANN map (dark blue), simulation of noise + CIB in the
  MILCA map (orange), simulation of noise+CIB in the MILCANN map
  (red).}
\label{distr}
\end{figure}

In Figure \ref{distr}, we compare the intensity distributions in MILCA
and MILCANN map.  For the MILCA map, we observe a significant tail of pixels with
negative intensity (mainly associated with radio-source contamination). We
do not observe this contamination in the MILCANN map, since the ANN
weighting process significantly reduces radio source contamination. We also
observe that the noise in the MILCANN map is lower than in the MILCA
map by a factor of five. However, we note that the intensity of the
brightest pixels in the map is not affected by the ANN weighting process.\\ 
Considering the
correlation between the ANN weight and the noise in MILCA map, the
noise in the MILCANN map does not present symmetric distribution. So,
we are dealing with a non-Gaussian, inhomogeneous, correlated noise
with an asymmetric distribution. As observed in Fig.\ref{distr}, the
noise is more likely to produce positive  than
negative values in the MILCANN map implying a non-zero expectation
value.  Consequently, in the following we used and propagated the
complete noise distribution.

\subsection{Noise inhomogeneities}
Due to the $Planck$ scanning strategy, the noise level on the sky is
inhomogeneous. The most noticeable feature is the fact that ecliptic poles present a higher redundancy of observations and thus a significantly lower noise level
\citep{PlanckMIS}. The MILCANN map, obtained from the product of the filtered MILCA map and the ANN weight, therefore exhibits noise
inhomogeneities amplified from the input noise in MILCA map.

\begin{figure}[!th]
\begin{center}
\includegraphics[scale=0.2]{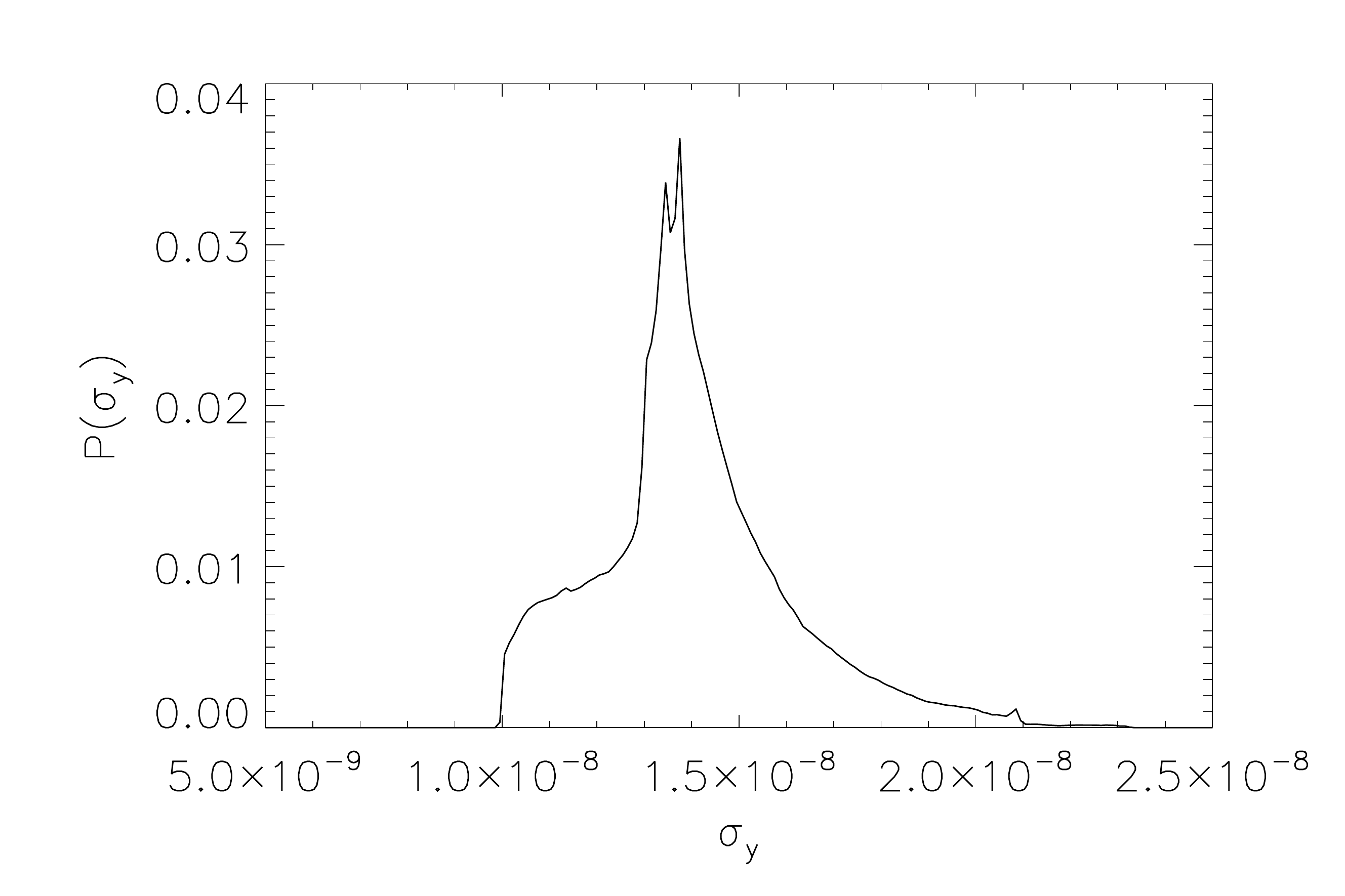}
\caption{Distribution of the noise standard deviation, $\sigma_y$ across the MILCANN full-sky map.}
\label{nhom}
\end{center}
\end{figure}

From the instrumental noise+CIB MILCANN simulated map, we derived the standard deviation of the noise in MILCANN map, $\sigma_y$, by computing the local standard deviation of
MILCANN simulated noise map within a four-degree Gaussian window. The  distribution of noise standard deviation, $\sigma_y$, is shown in Fig.~\ref{nhom}. It represents the distribution of the pixel-dependent noise levels.

\label{secnh}

We constructed a map of the signal-to-noise ratio, $\widehat{y}_\sigma$, as
\begin{align}
\widehat{y}_{\sigma} = \frac{\widehat{y}_{f}}{\sigma_{\rm y}}.
\end{align}
We note that using a unique threshold on $\widehat{y}_{\sigma}$
is equivalent to using a pixel-dependent threshold on
$\widehat{y}_{f}$.

\begin{figure}[!th]
\begin{center}
\includegraphics[scale=0.2]{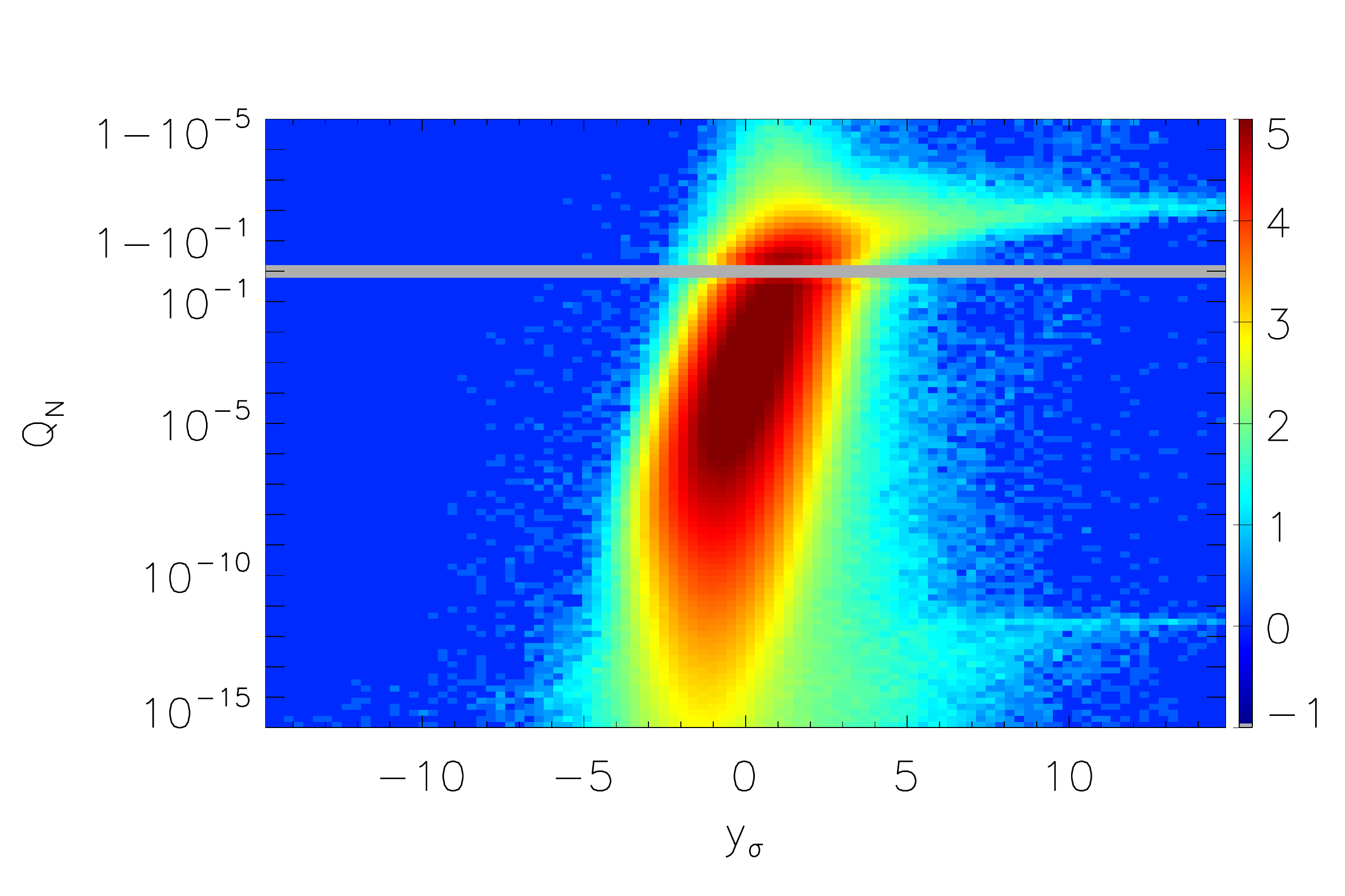}
\includegraphics[scale=0.2]{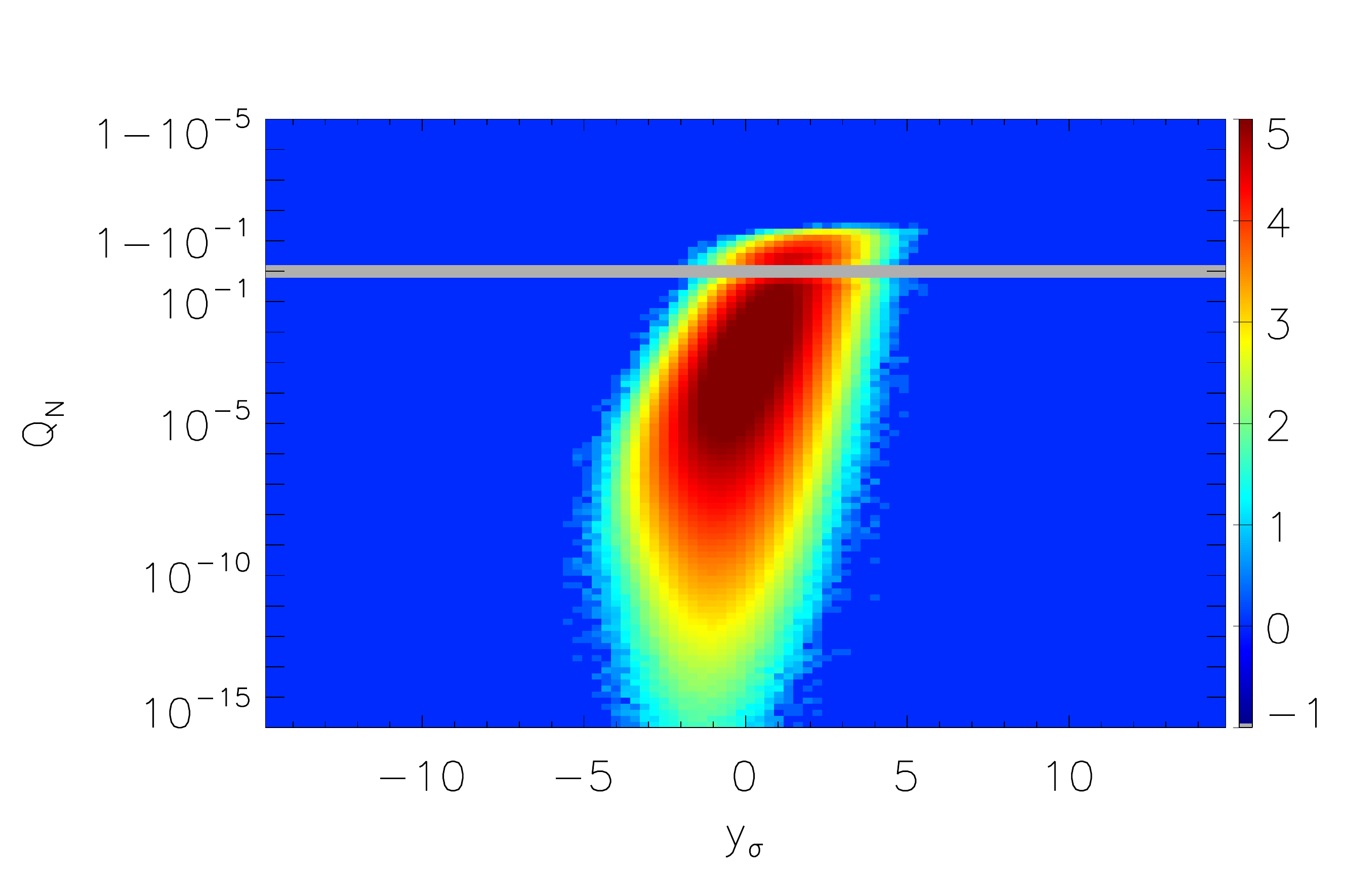}
\caption{Distribution of the MILCANN map (top panel) and MILCANN noise simulation (bottom panel) pixels as function of $Q_N$ and normalized intensity, $y_\sigma$. The colors show the number of pixels in logarithmic scale. To display the large dynamical range of the $y$ axis, we divided the axis into two log-scales with a "regular" log scale for small $Q_N$ values and "inverted" log-scale for large $Q_N$ values.}

\label{qnY}
\end{center}
\end{figure}

Figure~\ref{qnY} presents the distributions of MILCANN map and MILCANN
noise simulation as a function of $y_\sigma$ and $Q_{\rm
  N}$. We observe that the MILCANN noise simulation does not show
high-$Q_{\rm N}$ and high-$y_\sigma$ pixels ($Q_{\rm N} > 0.9$, that are associated with real tSZ signal). We also observe that for
$Q_{\rm N} \simeq 0$ the MILCANN map presents a significantly larger
distribution of $y_\sigma$ than the noise simulation (for $Q_{\rm N} \in [10^{-11},10^{-15}]$. This extended distribution is produced
by foreground residuals that are present in the MILCA tSZ map. We
  verified that these residuals are strongly correlated with the
  galactic latitude, implying that they are related to systematic
  effects. We do not observe a similar behavior at larger
values of $Q_{\rm N}$. This confirms that foreground residuals are
strongly reduced using the ANN weighting procedure, as already
observed on Fig.~\ref{distr}.

\section{tSZ candidate detection}
\label{carac_sec}

As shown by \citet{mel12}, tSZ-map based galaxy
   cluster detection methods suffer from a high level of contamination
   by spurious sources since it is difficult to
   disentangle real tSZ emission from biases in the tSZ map induced by residuals from
   astrophysical emissions. Given its significant reduced residual signals, the MILCANN map may be better
   suited for galaxy-cluster detection than standard reconstructed $y$ maps.
   In this section, we thus used the improved tSZ map obtained after
  applying the ANN-based weight to perform a basic cluster
  detection. We also characterize the purity and completeness of the
  derived sample of cluster candidates to assess the improvement
  compared to previous clusters catalogue derived from the $Planck$
  data. However, we stress that the MILCANN map cannot be used to
   provide accurate estimates of the flux or the shape of tSZ
   sources considering that the tSZ signal is affected by the ANN weighting response.

\subsection{Methodology}

To detect sources in MILCANN map, we applied a mask of the
galactic plane and point sources detected by $Planck$ keeping 84\% of
the sky as defined in \citet{planckPSZ}. Then, using the noise standard-deviation map computed in Sect.~\ref{sec_noise}, we divided the
MILCANN map by its local noise level map to perform the detection in signal-to-noise unit.
We applied a
threshold, $y_\sigma > 3$, to the MILCANN map, and considered as a candidate tSZ source adjacent pixels above the threshold. We discarded all sources detected on less than 5 adjacent pixels of 1.7x1.7
arcmin$^2$ to avoid detections produced by anomalous pixels. We cleaned
multiple detection of a same source by merging all sources in a
radius of 10 arcmin. We obtained a sample 3969 tSZ sources that we refer to as the HAD (Hurier, Aghanim and Douspis 2019) catalogue\footnote{The catalogue is available in the download section of \url{http://szcluster-db.ias.u-psud.fr}}.

\subsection{Characterization of the MILCANN detection method}

In this section, we present in
detail each step that allows us to compute the selection function of the HAD catalogue, that is a detailed description
of the transfer function of the galaxy-cluster signal across our processing.

\subsubsection{Fourier space filtering response}

\begin{figure}[!th]
\begin{center}
\includegraphics[scale=0.2]{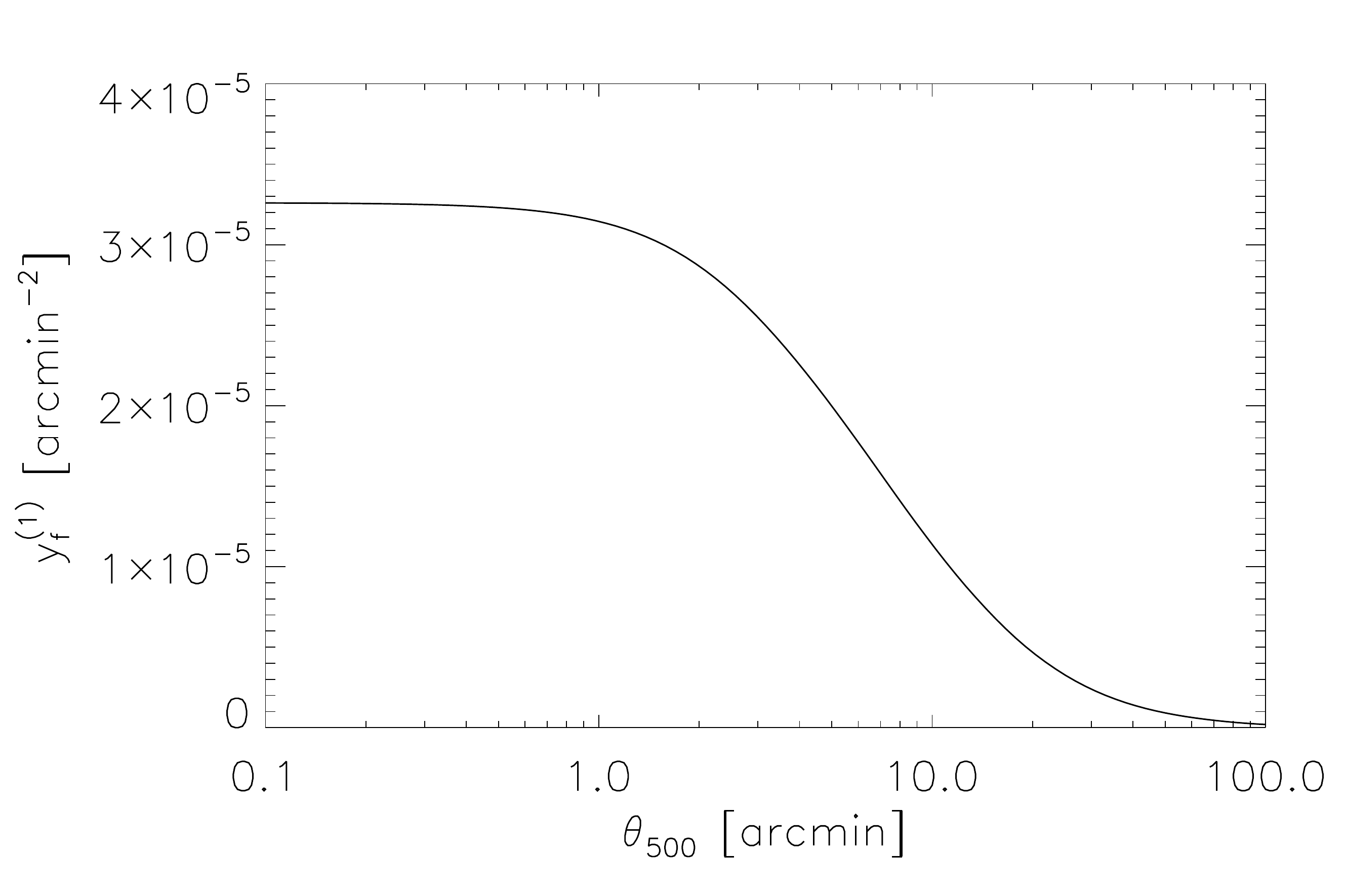}
\caption{Central intensity of the SZ signal in a filtered map for a galaxy cluster with
  $Y_{500} = 1$ arcmin$^2$ as a
  function of galaxy cluster typical radius $\theta_{500}$.}
\label{fresp}
\end{center}
\end{figure}
Before applying the ANN weight to build the MILCANN map, we have filtered the MILCA map with the matched-filter presented in Fig.~\ref{fl}.  Here, we present the
estimation of the transfer function of this filtering process.
To do so, we first build a mock map of a sky projected tSZ signal from a galaxy
cluster with $Y_{500} = 1$ arcmin$^2$ assuming a GNFW pressure profile
\citep{arn10} with 1000 pixels per $R_{500}$.  Then, we convolve
the tSZ mock map by the instrumental beam and by the matched-filter
presented in Sect.~\ref{ann_sec}.  We perform this procedure for
values of $R_{500}$ ranging from 0.1 to 100 arcmin.  Finally, we
extract the tSZ intensity at the center of the galaxy cluster on the
convolved mock map.\\ 
In Fig.~\ref{fresp}, we present the tSZ
intensity, $y^{(1)}_f$, after applying the matched-filter for a
galaxy cluster with a universal pressure profile and a flux $Y_{500} = 1$ arcmin$^2$ as a function of the apparent size on the sky, $\theta_{500}$. The matched-filter we
use selects compact objects (of typical size 5 arcmin) and thus presents a response that 
significantly reduces the flux of extended galaxy clusters. However, this is not an important limitation
  since our main goal is to detect compact tSZ sources associated with
  new galaxy clusters that are either low-mass or
  high-$z$. Considering the resolution of $Planck$ tSZ maps (roughly 7
  arcmin) such galaxy clusters are point-like. Furthermore for these
  clusters or for more extended ones, we can compute their tSZ signal
  direcly from the MILCA map or from the frequency maps.

\subsubsection{Completeness}

Given all the steps detailed above, we can express the complete processing we applied as
\begin{align}
y_\sigma =  Y_{500} \frac{ y^{(1)}_f (\theta_{500}) Q_{\rm N}(y_f)}{\sigma_y} + {N_\sigma},
\end{align}
where $Y_{500}$ is the tSZ flux of the galaxy cluster, $y^{(1)}_f$ is the
match-filtered central intensity for a cluster with $Y_{500} = 1$ arcmin$^2$, shown in
Fig.~\ref{fresp}, the dependency of $Q_{\rm N}$ with $y_f$ is
presented in Fig.~\ref{aresp}, the distribution of the noise across
the map, $\sigma_y$, is shown in Fig.~\ref{nhom}, and $N_\sigma$ is
the homogenized noise\footnote{That can be modelled through noise+CIB residuals simulations normalized by the standard deviation map $\sigma_y$} in the MILCANN map.\\
\begin{figure}[!th]
\begin{center}
\includegraphics[scale=0.2]{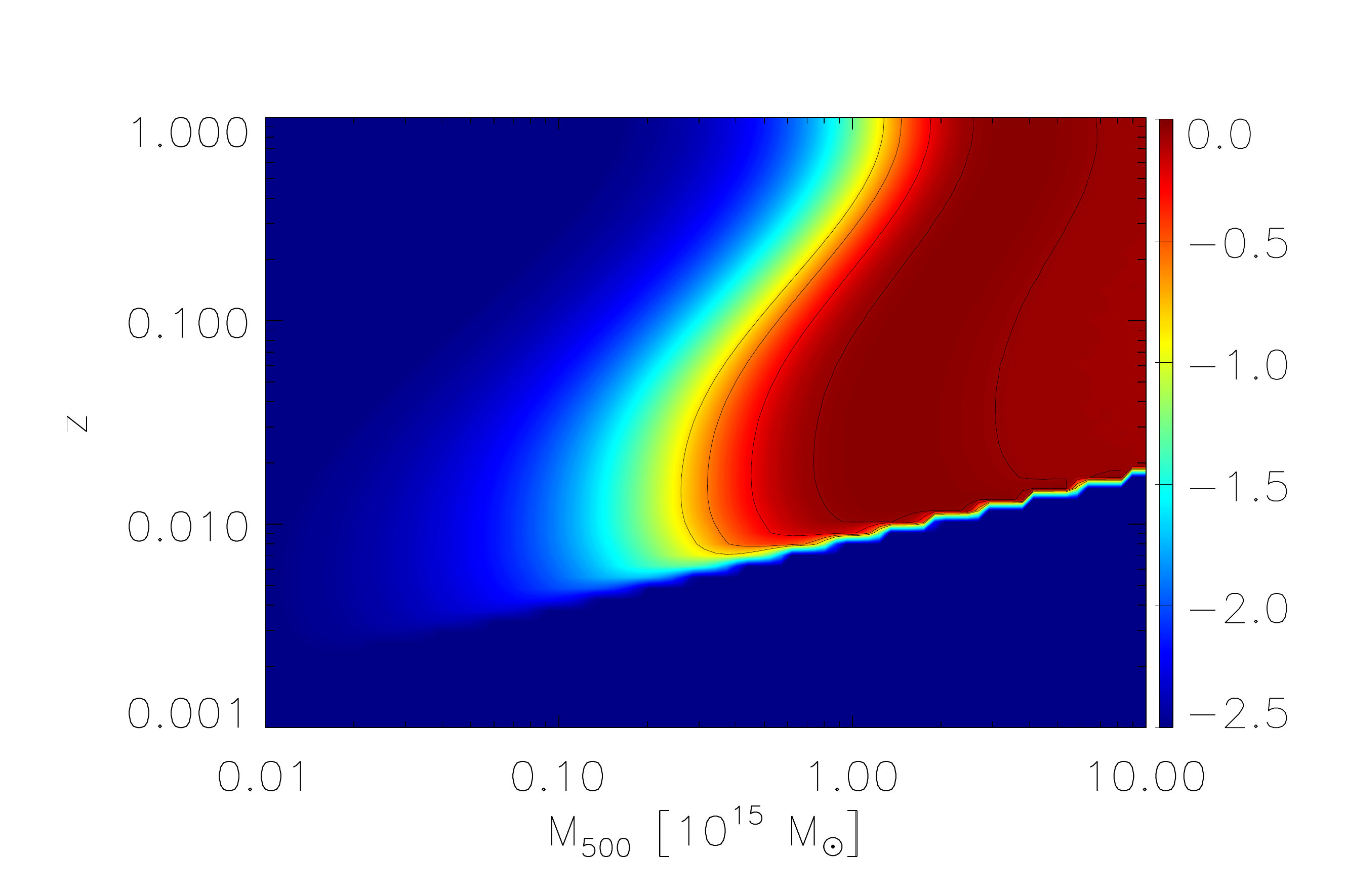}
\caption{Completeness of the HAD catalogue as a function of $M_{500}$ and
  $z$. The color scale is logarithmic and ranges from $10^{-2.5}$
  (dark blue) to $1$ (red). Black contours shows the 10, 20, 50, and 90\% completeness level.}
\label{selfunc}
\end{center}
\end{figure}

The galaxy cluster signal probability distribution is obtained by the convolution of: (i) the $M_{500}-Y_{500}$ relation intrinsic scatter, (ii) the distribution of $\sigma_y$ (noise inhomogeneity), and (iii) the distribution of the noise $N_\sigma$ (noise probability distribution). 
We assumed that the relation
  $M_{500}-\theta_{500}$ does not present any scatter.\\ The completeness,
${\cal C}(y_\sigma)$, is then given by the ratio of the integral of
$y_\sigma$ distribution, ${\cal P}(y_\sigma)$, above the detection
threshold normalized by the integral of the full distribution,
\begin{align}
{\cal C}(y_\sigma) = \frac{1}{\int_{0}^{\infty} {\cal P}(y_\sigma) {\rm d}y_\sigma}\int_{t}^{\infty} {\cal P}(y_\sigma) {\rm d}y_\sigma,
\end{align}
where $t$ is the detection threshold applied on the $y_\sigma$ map (3 in our case).

Figure~\ref{selfunc} shows the completeness as a function of the mass,
$M_{500}$ and redshift, $z$, of a given cluster. We observe that, with a very basic detection method applied on a filtered and cleaned tSZ map, we
can detect
clusters down to a typical mass of $M_{500} = 1\, 10^{14}$
$M_\odot$ with percent level completeness. We also observe that for very large mass ($ > 2\, 10^{15}$
$M_\odot$) the completeness is slightly smaller than one. This effect is
produced by the matched-filter that significantly reduces the tSZ effect
produced by extended (massive) sources.

\subsubsection{Purity}

We estimated the purity of the catalogue by performing the
detection of tSZ sources on the MILCANN map and the MILCANN noise
simulation from Sect.~\ref{sec_noise}. This estimate does not
account for all foreground residuals in the MILCANN map which thus
may slightly over-estimate the purity of the HAD catalogue.  We
found $N^{(1)}_{\rm det}$ detections for the MILCANN map and
$N^{(2)}_{\rm det}$ for the simulated noise map. We performed the
detection for several detection thresholds.  The purity is obtained as
$P = \frac{N^{(1)}_{\rm det} - N^{(2)}_{\rm det}}{N^{(1)}_{\rm det}}$.
\begin{figure}[!th]
\begin{center}
\includegraphics[scale=0.2]{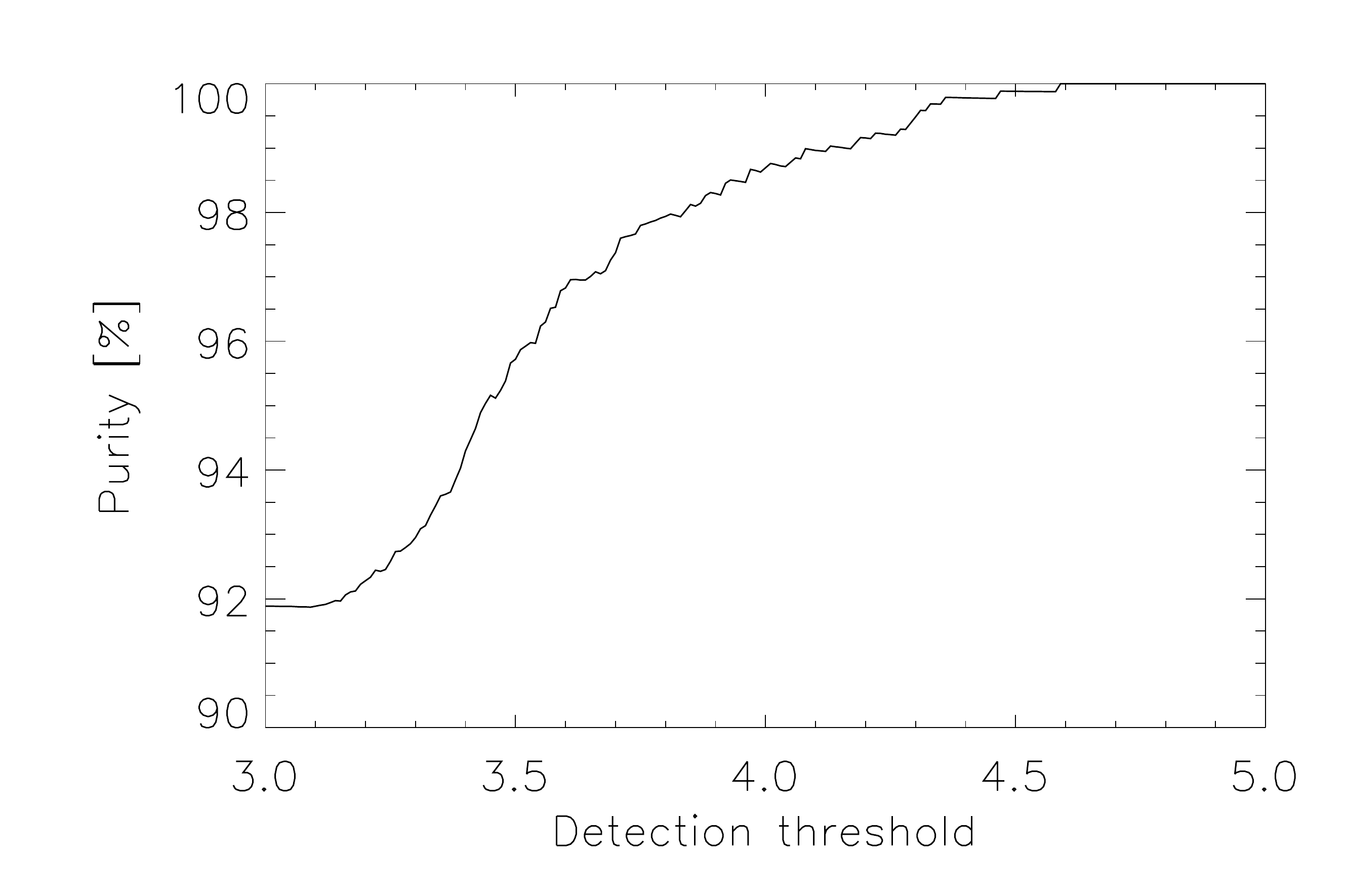}
\caption{Purity of the HAD catalogue as a function of the detection threshold.}
\label{purefig}
\end{center}
\end{figure}
In figure ~\ref{purefig}, we present the purity as a function of the
detection threshold. For the threshold, $D_t = 3$ , used in the construction of the HAD catalogue, we derived an estimated purity above 90\%.

\subsection{Comparison with reference galaxy cluster catalogues}

In this section, we present a brief comparison of the HAD
cluster candidate catalogue and other reference catalogues.  We
  used two approaches for the comparison.\\ First, we compared the
  numbers of cluster candidates in the HAD catalogue, with the
  predicted number of clusters assuming $Planck$-SZ cosmology
  \citep{planckSZC} considering the completeness and purity of the
  HAD catalogue (see above). This predicted number is found to be $4082 \pm 700$ (the uncertainty is obtained by propagating the uncertainty over $\Omega_{\rm m}$ and $\sigma_8$ to galaxy cluster number count).
  Thus, the 3969 detected candidates in the HAD catalogue is
  consistent with this prediction. 
 However,  this number has to be considered carefully as the assumptions on the scaling relation and completeness may not encompass the full complexity of the cluster physics.

\begin{figure}[!th]
\begin{center}
\includegraphics[scale=0.2]{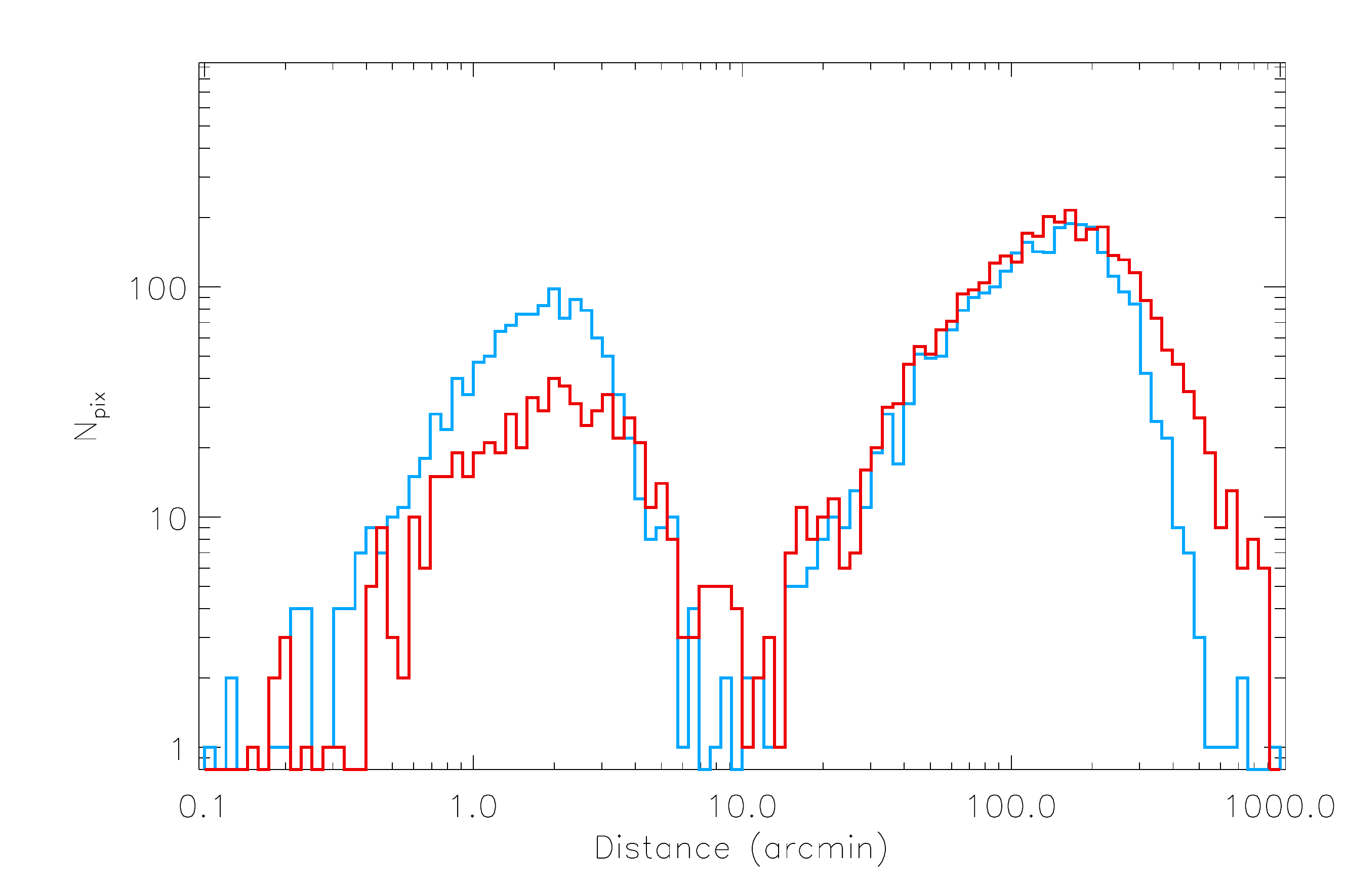}
\caption{Nearest neighbor distance distribution between HAD sources and PSZ2 public catalogue sources (Blue histogram) and MCXC public catalogue sources (Red histogram).}
\label{psz2}
\end{center}
\end{figure}

Then, we performed a cross-match with reference galaxy cluster catalogues.  We compared our catalogue of candidates with the
  PSZ2 catalogue. The distribution of distance to nearest neighbor is shown in Fig.~\ref{psz2}
 Among the 3969 HAD sources, we find 1243 in common with
  the PSZ2 sources among which 997 are confirmed galaxy clusters all having an ANN quality factor $Q_{\rm Neural} > 0.6$ as defined in
    \citet{agh14}.  The HAD catalogue additionally contains 496 known
  clusters that are missing from PSZ2 (they are contained in the ACT, SPT,
  redMaPPer, Wen+12, or MCXC catalogues). Figure~\ref{psz2} shows
  clearly the two populations of sources common and not common with PSZ2. For the
  sources common to HAD and to PSZ2 the typical separation is below 4 arcmin consistent with the $Planck$ resolution.  For very extended sources the position mismatch between HAD and PSZ2 can reach up to 10 arcmin. It strongly depends on the
  exact methodology used to define the galaxy cluster
  position. Nevertheless, considering the number density in $Planck$ SZ
  catalogue and HAD catalogue, within a radius of 10 arcmin the number of
  chance association is 15 ($\simeq 1\%$ of the PSZ2 sample) and within a radius of 4
  arcmin this number is 4. It implies that a 10 arcmin matching distance still provides a robust association between HAD and PSZ2 sources.\\
  Figure~\ref{psz2} presents the distribution of distance nearest neighbors between HAD and MCXC objects.
 The matching procedure of the HAD catalogue outputs the following positional associations with reference catalogues:
\begin{itemize}
\item[$\bullet$] 1243 HAD sources in common with PSZ2
\item[$\bullet$] 601 HAD sources matched with known X-ray clusters from MCXC
  (including 92 objects not in PSZ2)
\item[$\bullet$] 687 HAD sources matched with over-density of galaxies ($N_{200}$ > 25) from
  WHL12 (including 276 objects not in PSZ2), we estimated
  a maximum number of chance association of 20 at 99\% confidence
  level.
  \item[$\bullet$] 115 HAD sources matched with over-density of galaxies ($N_{500}$ > 8) from
  new clusters of WHL15 (including 79 objects not in PSZ2), we estimated
  a maximum number of chance association of 18 at 99\% confidence
  level.
 \item[$\bullet$] 1400 HAD sources matched with over-density of galaxies ($R_{L}$ > 10) from
  WHY18 (including 649 objects not in PSZ2), we estimated
  a maximum number of chance association of 50 at 99\% confidence
  level.
\item[$\bullet$] 469 HAD sources matched with over-density of galaxies ($\Lambda$
  > 50) from redMaPPer (including 179 objects not in PSZ2)
\item[$\bullet$] 35 HAD sources matched with SPT clusters
\item[$\bullet$] 43 HAD sources matched with ACT clusters
\end{itemize}
The cross-matched numbers are summarized in Table.~\ref{tab_cross}. Objects present in the PSZ2 catalogue that are not present in the HAD catalogue (418 sources in total) consists of extended sources (smeared out by our matched-filter) or sources with a very low quality flag from the ANN. A low quality ANN quality assessment implies either that these sources are spurious detection or are showing contamination by at least another type of astrophysical emission.

\begin{table*}
\center
\caption{Cross-matched number of objects between HAD and reference catalogues. We also performed the same cross-match for the PSZ2. The total column refers to the number objects from HAD or PSZ2. For Wen+12 catalogue we imposed $N_{200} > 25$, for Wen+15 we imposed $N_{500} > 8$ and for redMaPPer we imposed $\lambda > 50$.}
\begin{tabular}{c|ccccccccc|c}
\label{tab_cross}
& HAD & PSZ2 & SPT & ACT & MCXC & WHL12 & WHL15 & WHY18 & redMaPPer & TOTAL\\
\hline
N$_{\rm obj}$ & 3969 & 1653 & 224 & 182 & 1743 & 9951 & 8625 & 47594 & 5540 & \\
HAD & / & 1243 & \phantom{0}35 & \phantom{0}43 & \phantom{0}601 & \phantom{0}687 & \phantom{0}115 & \phantom{0}1400 & \phantom{0}469 & 1803 \\
PSZ2 & 1235 & / & \phantom{0}27 & \phantom{0}30 & \phantom{0}556 & \phantom{0}427 & \phantom{00}48 & \phantom{00}881 & \phantom{0}312 & 1134 \\

\hline
\end{tabular}
\end{table*}

 We used the redMaPPer red-sequence based redshifts to compute the
  redshift and richness, $\Lambda$, distributions of the matching
  population between HAD and redMaPPer catalogues in two cases: (i)
  considering all cluster candidates in the HAD catalogue that
  match redMaPPer sources and (ii) candidates in the HAD catalogue not
  contained in PSZ2 catalogue that match redMaPPer sources.

\begin{figure}[!th]
\begin{center}
\includegraphics[scale=0.2]{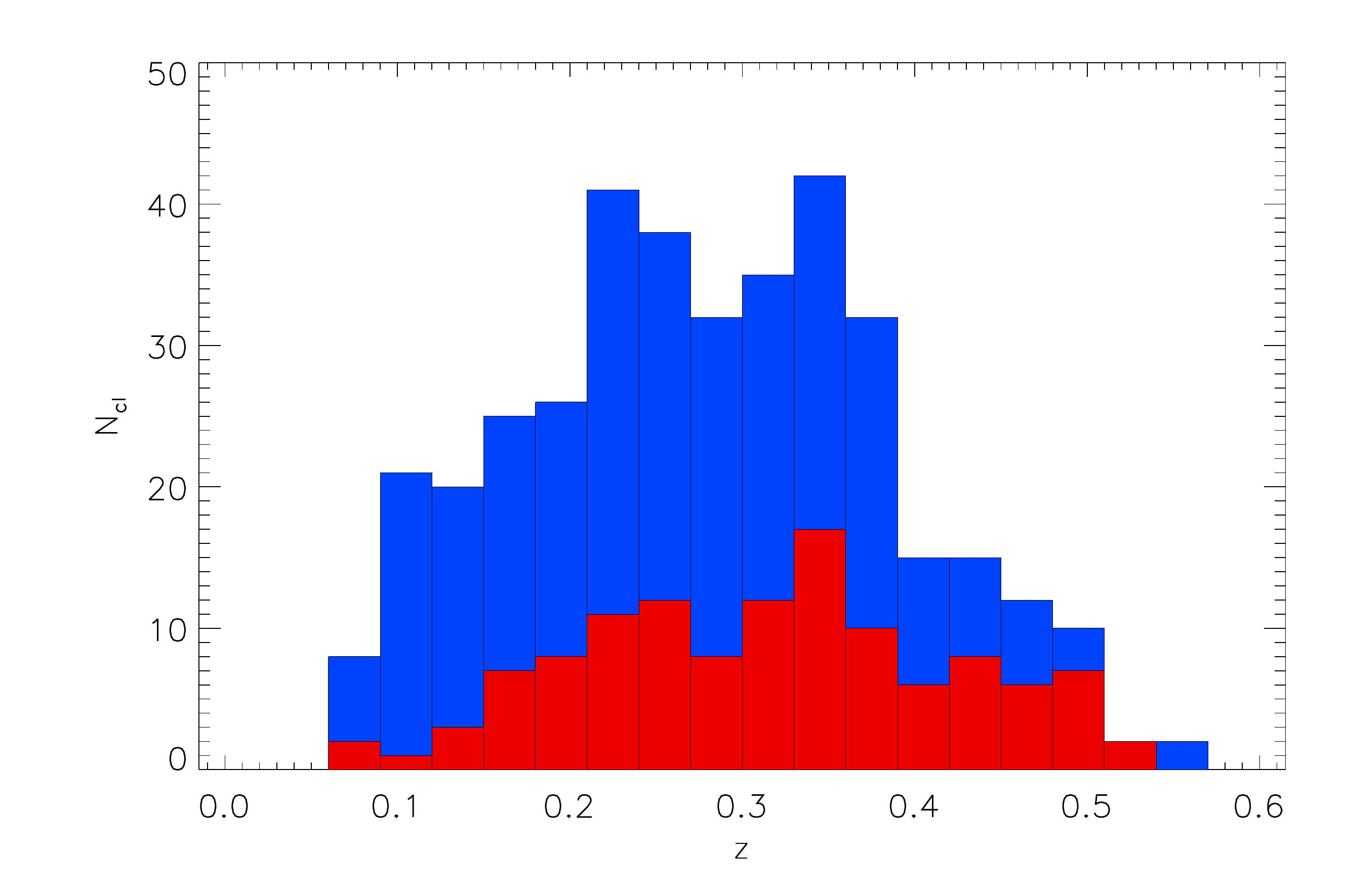}
\includegraphics[scale=0.2]{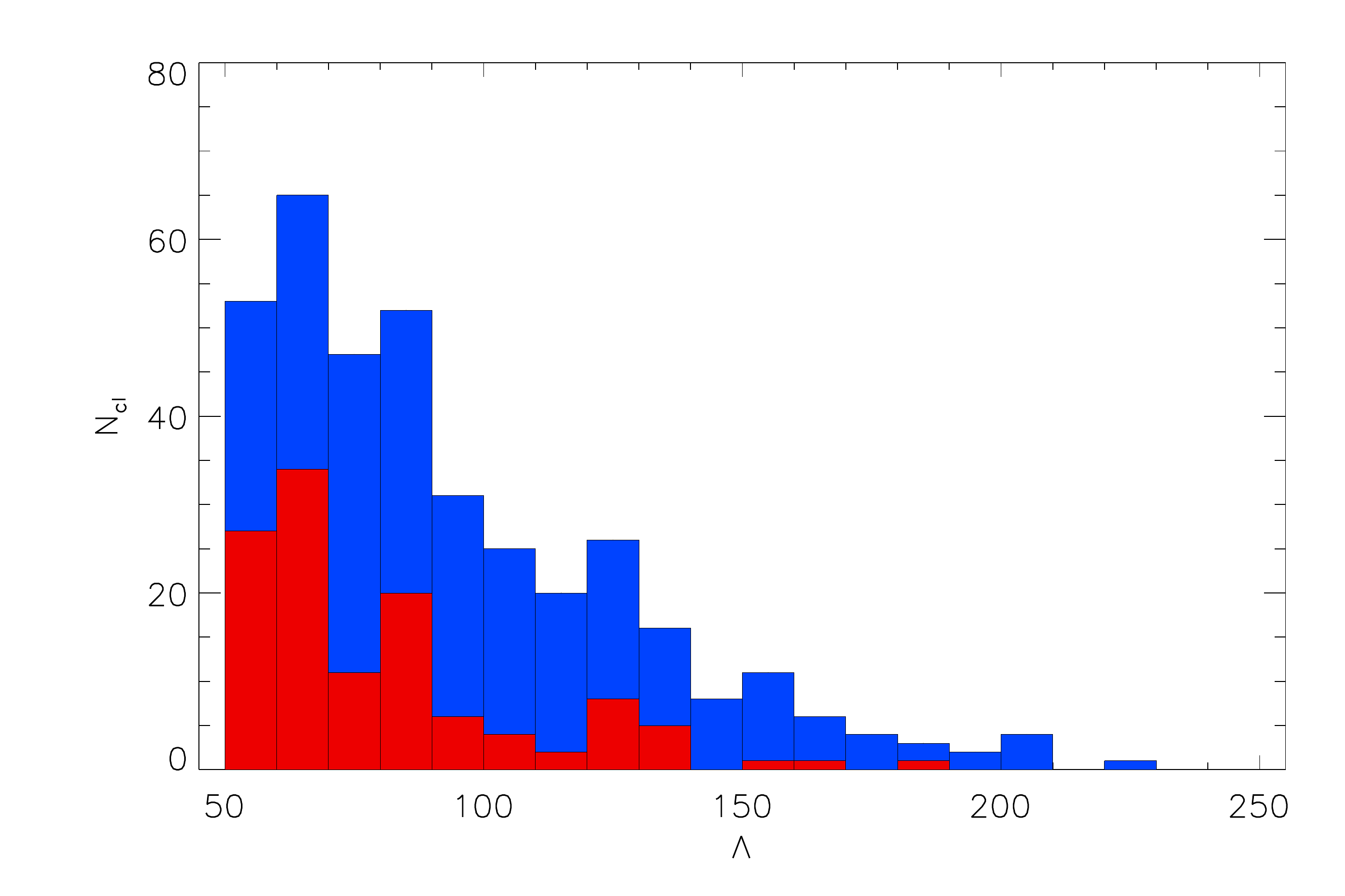}
\caption{Top panel: redsfhit distribution for HAD candidates matching
  redMaPPer overdensities of galaxies (dark blue) and for HAD
  candidates not contained in PSZ2 catalogue (red). Bottom panel:
  richness, $\Lambda$, distribution for HAD candidates matching
  redMaPPer overdensities of galaxies (dark blue) and for HAD
  candidates not contained in PSZ2 catalogue (red)}
\label{zdis}
\end{center}
\end{figure}

In figure ~\ref{zdis}, we present the redshift distribution for
sources contained both in HAD and redMaPPer catalogues.  We observe that
new tSZ sources in the HAD catalogue, that is not contained in the PSZ2
catalogue, are essentially at higher redshift than PSZ2
clusters. This redshifts distribution confirms that the MILCANN tSZ
map enables the detection of high-$z$ galaxy clusters that are not
contained in the PSZ2 catalogue. From this redshift distribution, we
find that new HAD sources are at redshifts ranging from 0.2 to
0.5. Figure~\ref{zdis} also shows the richness distribution for all
HAD sources and HAD sources not contained in the PSZ2 that match
redMaPPer galaxy overdensities. We observe that the HAD sources not
contained in the PSZ2 catalogue are preferentially at lower
richness.

\section{Multi-wavelength statistical characterization of The HAD sample}
\label{sec_probe}

In this section, we perform a stacking analysis to unveil the average
properties of the HAD sources. In particular, we
focus on the average sub-millimeter SED, on the stacked lensing
signal, and on color-color diagnostic using the WISE galaxy catalogue.

\subsection{Stacked SED of cluster candidates}

\begin{figure}[!th]
\begin{center}
\hspace{-1cm} \includegraphics[scale=0.2]{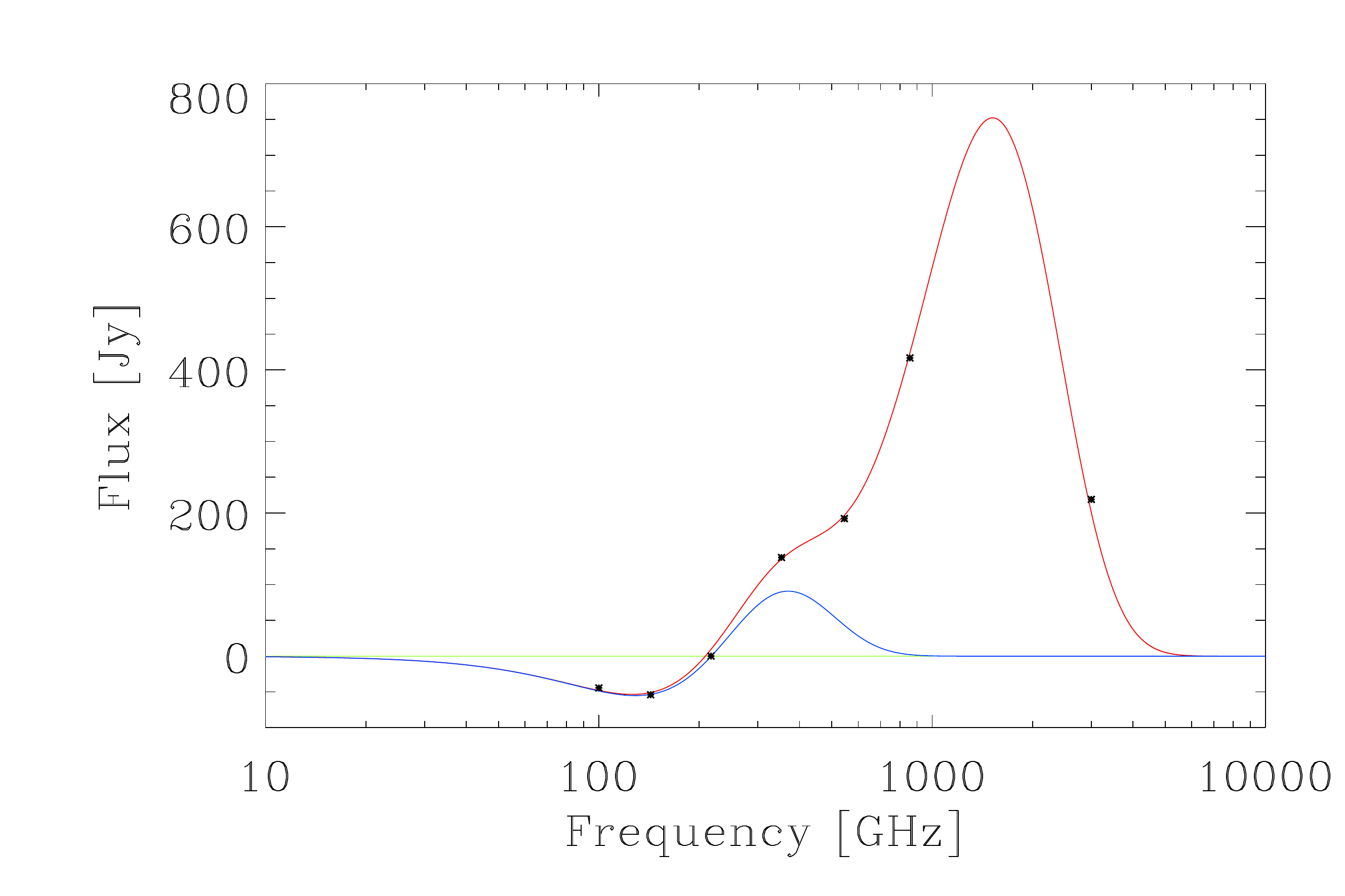}
\caption{Stacked SED of galaxy-cluster candidates (black sample). The
  tSZ effect contribution is shown as a solid blue line, and the total
  SED accounting for tSZ and infra-red emission as a solid red line.}
\label{stacksed}
\end{center}
\end{figure}

For cluster candidates in HAD not seen in PSZ2, we stack the $Planck$
maps per frequency and the IRIS full-sky map at 100 $\mu$m
\citep{miv05}. Then, we measure the flux through aperture photometry.
Figure ~\ref{stacksed} presents the obtained SED exhibiting both
tSZ and IR emissions. Consistently with \citet{planckszcib}, we modeled
the IR emission with a modified black body SED assuming
$\beta_d = 1.75$, a dust temperature $T_d = 24$ K, and a mean redshift
$\bar{z} = 0.4$. Comparing with previous analysis \citep{planckszcib},
the amount of IR emission toward HAD new tSZ sources is compatible with IR emission observed
toward PSZ2 confirmed galaxy clusters. We observed that the IR contribution is negligible at 100 and 143 GHz.
At 353 and 545 GHz, the IR emission contributes for 25 and 80\% of the total signal respectively.
At higher frequency, the tSZ contribution is negligible.

\subsection{CMB Lensing}

\begin{figure}[!th]
\begin{center}
\includegraphics[scale=0.25]{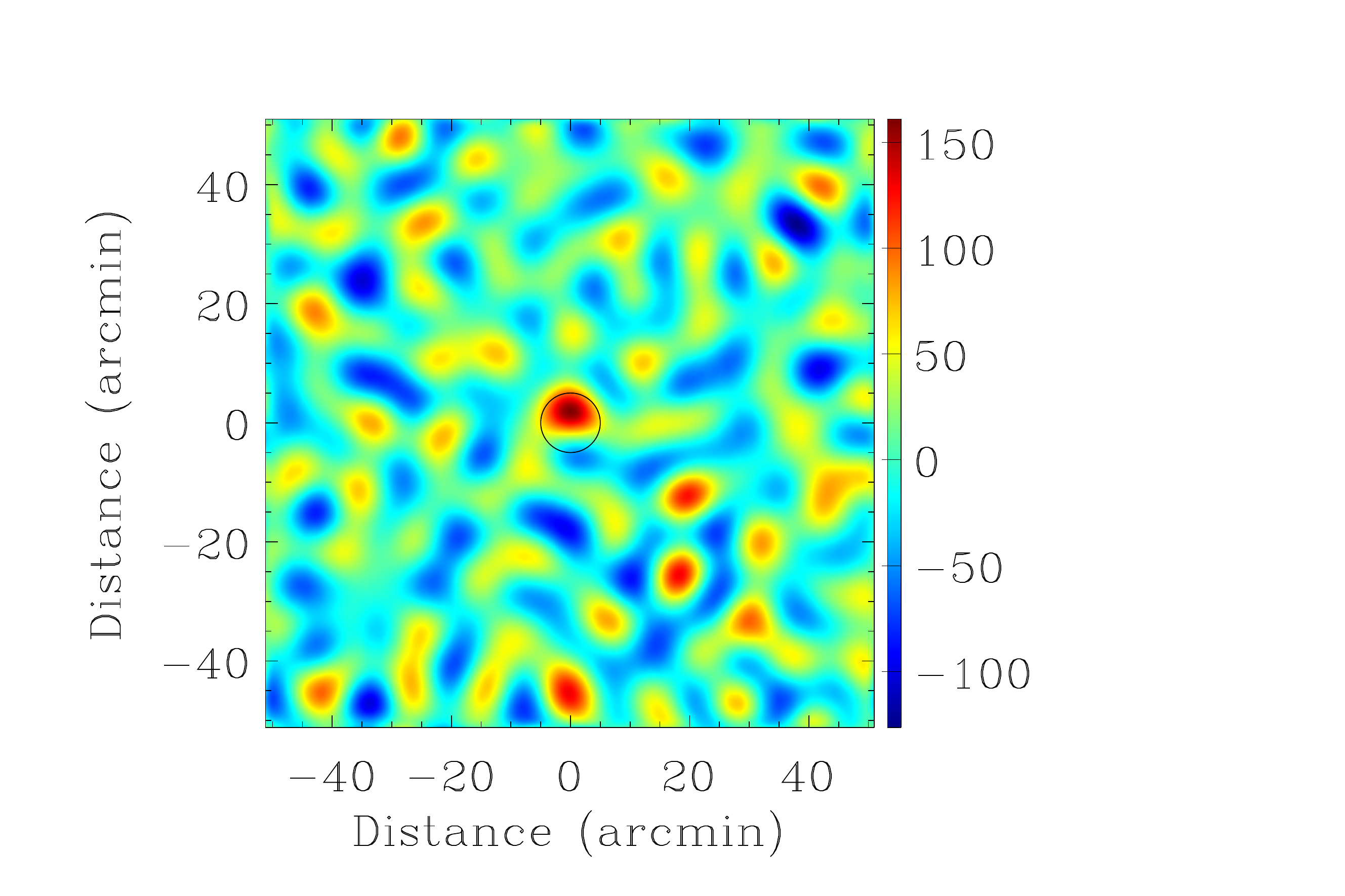}
\caption{Stacked CMB-lensing convergence map toward HAD SZ-candidates not contained in the PSZ2 catalogue.}
\label{stacklens}
\end{center}
\end{figure}

We stacked the CMB lensing convergence measured by the $Planck$ collaboration
\citep{plcklens} for all sources from the HAD catalogue that are not
included in the PSZ2 catalogue. In Fig.~\ref{stacklens},
we present the derived stacked signal.  We detect an excess in
convergence at a 5 $\sigma$ confidence level. Assuming a typical
redshift in the range 0.2 to 0.5 for galaxy clusters, and correcting
for purity we compute an average mass of about $2\, 10^{14}$ M$_\odot$
per source \citep[see e.g.,][for a detailed description of the
  convergence to mass conversion]{mel15}.

\subsection{WISE catalog}

\begin{figure}[!th]
\begin{center}
\includegraphics[scale=0.25]{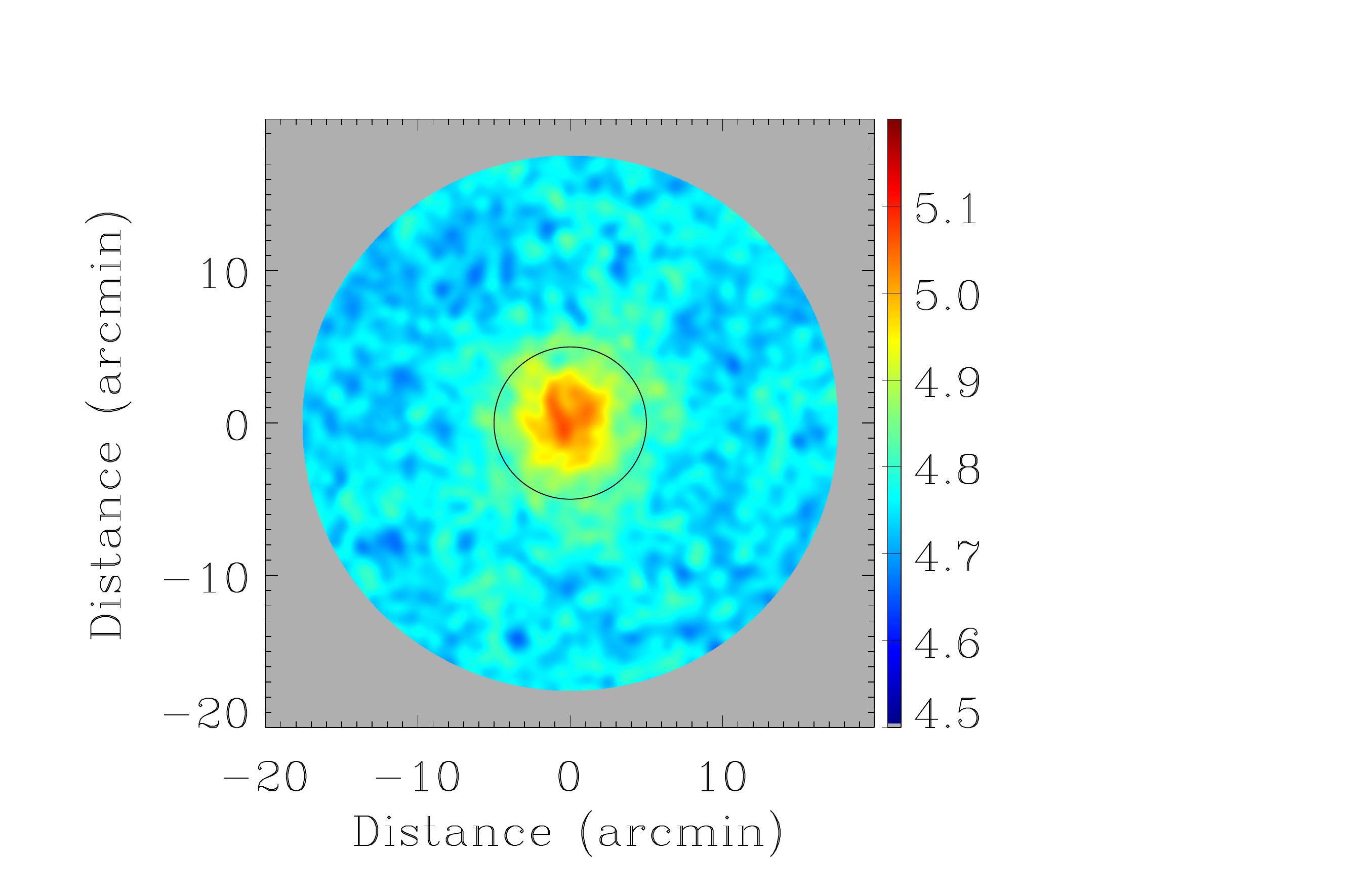}
\caption{Stacked number of sources in AllWISE catalogue toward SZ-candidates not contained in the PSZ2 catalogue.}
\label{stackwise}
\end{center}
\end{figure}

We also searched for counterparts of the HAD SZ-candidates in the AllWISE
full-sky catalogue \citep{cut13}.  We stacked the sources in the
AllWISE catalogue toward HAD sources that have no PSZ2
counterpart.  The stacked density of sources is presented in
Fig.~\ref{stackwise}. We observe a significant AllWISE source overdensity of $23 \pm 1$\footnote{The uncertainty provided is on the average over-density of sources and do not account for intrinsic scatter between sources in the stacked sample} per HAD source inside an aperture of 10
arcmin; the background level of the AllWISE source-density map was estimated between 10 and 15 arcmin.

\begin{figure}[!th]
\begin{center}
\includegraphics[scale=0.25]{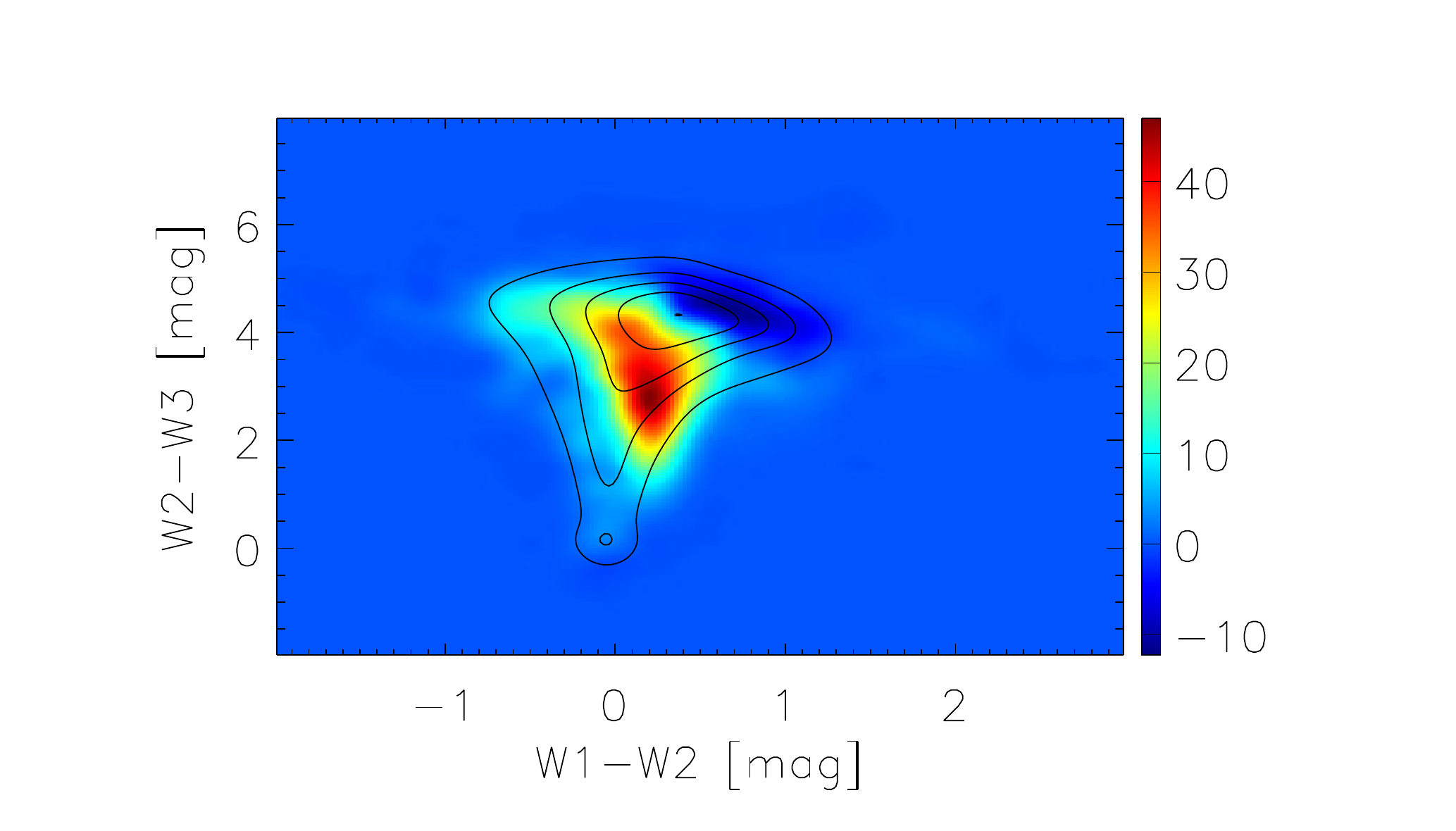}
\includegraphics[scale=0.25]{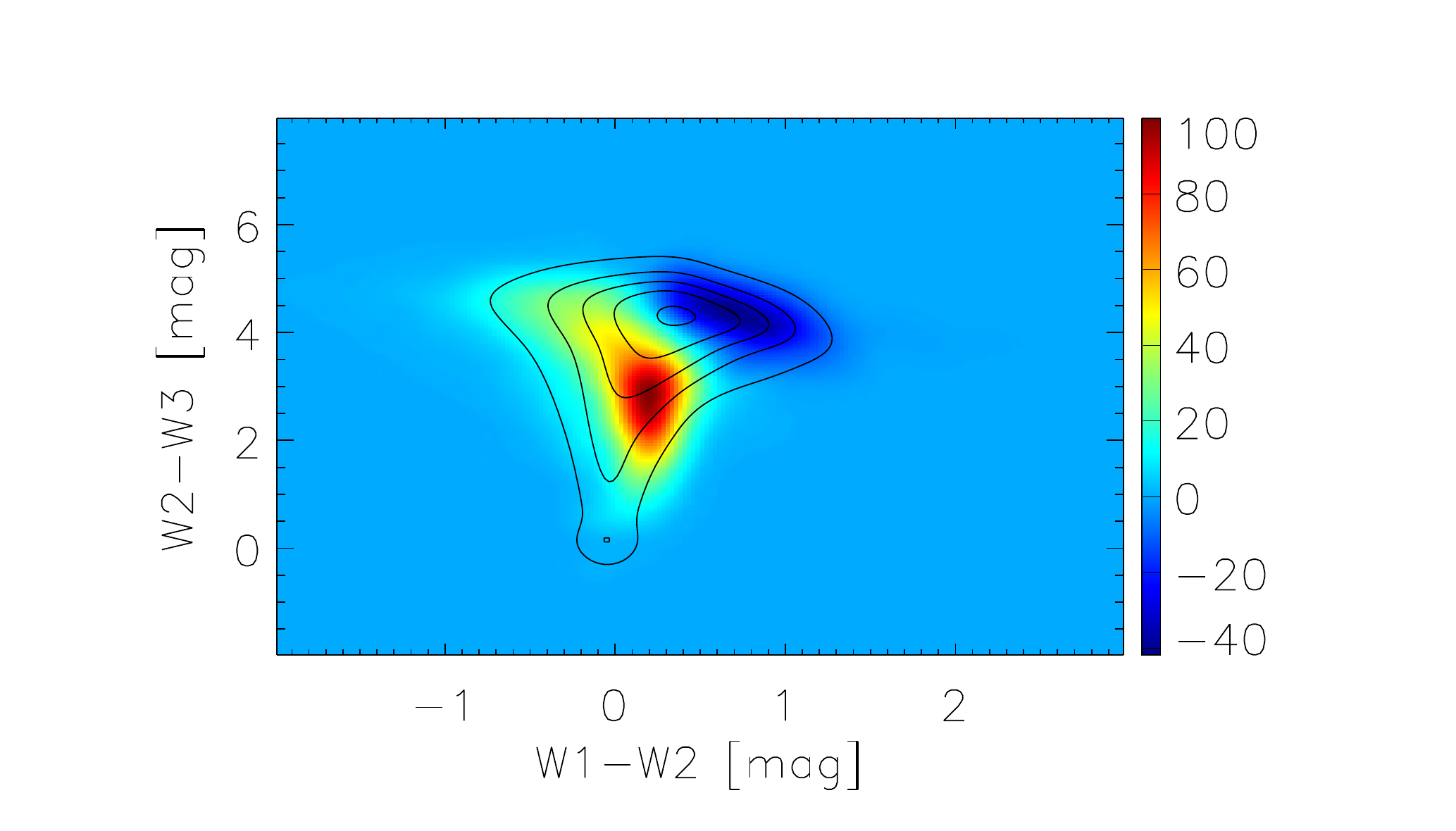}
\caption{WISE color-color plane for HAD SZ sources not contained (top panels) and contained (bottom panels) in the PSZ2. The color scale shows the distribution for member galaxies, for comparison the distribution for background object is represented by the black contours.}
\label{wcch}
\end{center}
\end{figure}

We also studied the distribution of the AllWISE matching sources in
the AllWISE color-color plane.  The surface density of HAD-source
member galaxies in the AllWISE catalogue is small compared to the total
density of all objects. Consequently, we cannot estimate the AllWISE
colors for each member galaxy individually and we 
estimated the color-color distribution of the cluster-member
galaxies.\\ First we compute the color-color distribution of AllWISE
sources, $D_{\rm in}$, within a radius of 10 arcmin around the HAD sources. Then, we
compute the same distribution, $D_{\rm out}$, for sources located between 10 and 15
arcmin from the HAD sources. Assuming that the background and
foreground objects are uniformly distributed on a 15 arcmin scale, we
estimated the distribution for member galaxies by computing the
sky-area weighted difference between the two distributions,
 \begin{align}
D_{\rm cl} = D_{\rm in} - \frac{A_{\rm in}}{A_{\rm out}}D_{\rm out},    
\end{align}
where $D_{\rm cl}$ is the color-color distribution of galaxy cluster members, $A_{\rm in}$ and $A_{\rm out}$ correspond to the area of the regions used to compute $D_{\rm in}$ and $D_{\rm out}$.\\

In Fig.~\ref{wcch}, we present the color-color distribution of member
galaxies of HAD sources not included in the PSZ2 catalogue and of HAD sources included the PSZ2 catalogue. We observe that both populations present
similar distributions in the (W2-W3)-(W1-W2) color-color plane. These
distributions are significantly different from the distribution of
background in the WISE catalogue.  They show a positive excess
for $W1-W2 \simeq 0.2$ and a lack of object for $W2-W1 \simeq 0.8$.\\

We used the SWIRE galaxy template library \citep{pol07}, and we
compute the expected tracks in the WISE color-color plane for 25
galaxy SED templates.  When comparing with tracks for various galaxy
types, we observe that the negative decrement in the WISE color-color
plane is essentially populated by AGNs or high redshift IR
sources. This implies that tSZ-based catalogues present a selection bias
toward clusters not hosting bright AGN or that present a strong
IR source in the foreground/background. Indeed, for clusters that
host a bright radio-loud AGN the tSZ effect can not be recovered in
the microwave and sub-millimeter domain. A similar argument applies
when the tSZ effect from a galaxy cluster is contaminated by bright
IR sources. From these color-color distributions, we can
conclude that the cluster candidates in the HAD catalogue are
populated by low-redshift ($z < 1$) elliptical, spiral galaxies, and
LRG as expected for tSZ samples derived with the $Planck$ experiment.

\section{Conclusion}

Previous studies have shown that a tSZ-map based approach is not optimal
  and less-efficient than a multi-frequency based approach
  \citep{mel12} to detect clusters of galaxies via their tSZ signal.  
  However, we have demonstrated that an ANN quality assessed tSZ map, MILCANN, enables
  to construct a competitive tSZ source catalogue even with a simple
  detection method. 
  
  The matched-filtering and the ANN weighting process involved in the construction of the MILCANN tSZ-map makes its utilization tailored for tSZ-cluster detection. In
  particular, the MILCANN tSZ-map presents a significantly lower level
  of noise and foreground residuals than standard tSZ maps.  However,
  the ANN weighting procedure produces a distorsion of the tSZ signal both in
  shape and in flux. Consequently, the MILCANN map can only be used for
  cluster detection purposes and it is not suited for other
  analysis such as tSZ scaling relations, profiles, or angular power
  spectrum.\\ From the MILCANN tSZ map, we have constructed the HAD
  source catalogue containing 3969 cluster candidates, with an
  estimated purity of 90\%. This catalogue is more than twice as large as the $Planck$ catalogue \citep{PSZ2}, achieving cluster detection down to
  $M_{500} = 10^{14}$ M$_\odot$, and reaches the same purity level.\\ We have verified that the number of sources in the HAD
    catalogue is consistent with the expected cluster abundance.
  \\ Additionally, comparing the HAD catalogue with ancillary catalogues, we
  demonstrated that the HAD galaxy clusters candidates contains new tSZ detections at high redshift and low richness. \\ 
  Finally we have shown that the sources detected in the
  MILCANN map present an excess of convergence in the {\it $Planck$} CMB lensing map,
  compatible with cluster masses of $2\,10^{14}$ M$_\odot$,
  and host galaxies with the same spectral behavior than $Planck$ PSZ2
  galaxy cluster member galaxies (elliptical, spiral galaxies, and LRG
  at $z < 1$).\\

\begin{acknowledgements}
The authors thank an anonymous referee for the comments that improved the publication. They thank the International space science institute (ISSI) for hosting initial discussions during the meetings of the
international team meeting "SZ effect in the $Planck$ era". GH acknowledges the support of the Agence Nationale de la Recherche
through grant ANR-11-BS56-015. This publication used observations obtained with \emph{Planck} (\url{http://www.esa.int/Planck}), an ESA science mission with instruments and contributions directly funded by ESA Member States, NASA, and Canada. It made use of the SZ-Cluster Database (http://szcluster-db.ias.u-psud.fr/sitools/client-user/SZCLUSTER\_DATABASE/project-index.html) operated by the Integrated Data and Operation Centre (IDOC) at the Institut d'Astrophysique Spatiale (IAS) under contract with CNES and CNRS. This project has received funding from the European Research Council (ERC) under the European Union's Horizon 2020 research and innovation programme grant agreement ERC-2015-AdG 695561.
\end{acknowledgements}

\bibliographystyle{aa}
\bibliography{milcann.bib}

\end{document}